\documentclass[aip,twocolumn,reprint,groupedaddress]{revtex4-1}

\usepackage{color,graphicx}
\usepackage{amsmath}
\usepackage{natbib}
\usepackage{pifont}
\usepackage{chemformula}
\usepackage{siunitx}

\newcommand{\bfo}{BiFeO$_3$}
\renewcommand{\selectlanguage}[1]{}

\begin{document}

\title{Giant Strain Tunability in Polycrystalline Ceramic Films via Helium Implantation}

\author{A.~Blázquez~Martínez$^{1,2,3}$}
\author{S.~Glin\v{s}ek$^{1,3}$}
\author{T.~Granzow$^{1,3}$}
\author{J.-N.~Audinot$^1$}
\author{P.~Fertey$^4$}
\author{J.~Kreisel$^{2,3}$}
\author{M.~Guennou$^{2,3}$}
\author{C. Toulouse$^{1,2,3,5}$}
\email{constance.toulouse@cnrs.fr}

\affiliation{$^1$Smart Materials Unit, Luxembourg Institute of Science and Technology, 41 rue du Brill, 4422 Belvaux, Luxembourg\\
$^2$Department of Physics and Materials Science, University of Luxembourg, 41 rue du Brill, 4422 Belvaux, Luxembourg\\
$^3$Inter-institutional Research Group Uni.lu–LIST on ferroic materials, 41 rue du Brill, 4422 Belvaux, Luxembourg\\
$^4$Synchrotron SOLEIL, L'Orme des merisiers, Saint-Aubin, Gif-sur-Yvette, France\\
$^5$CRISMAT Laboratory, University of Caen, CNRS UMR-6508, ENSICAEN, 6 Bd du Maréchal Juin, F-14000 Caen, France}


\begin{abstract}

Strain engineering is a powerful tool routinely used to control and enhance properties such as ferroelectricity, magnetic ordering, or metal-insulator transitions. Epitaxial strain in thin films allows manipulation of in-plane lattice parameters, achieving strain values generally up to 4\%, and even above in some specific cases. In polycrystalline films, which are more suitable for functional applications due to their lower fabrication costs, strains above 1\% often cause cracking. This poses challenges for functional property tuning by strain engineering. Helium implantation has been shown to induce negative pressure through interstitial implantation, which increases the unit cell volume and allows for continuous strain tuning with the implanted dose in epitaxial monocrystalline films.  However, there has been no study on the transferability of helium implantation as a strain-engineering technique to polycrystalline films. Here, we demonstrate the technique's applicability for strain engineering beyond epitaxial monocrystalline samples. Helium implantation can trigger an unprecedented lattice parameter expansion up to 3.2\% in polycrystalline \bfo{} films without causing structural cracks. The film maintains stable ferroelectric properties with doses up to \SI{E15}{He.cm^{-2}}. This finding underscores the potential of helium implantation in strain engineering polycrystalline materials, enabling cost-effective and versatile applications.
\end{abstract}

\maketitle

\section{Introduction}

Tuning and controlling functional properties is a key challenge for materials sciences and its applications. A powerful way to control, enhance, or even induce functional properties that is now routinely implemented, is the use of strain-engineering. It allows the tuning of ferroelectric properties\cite{Choi2004, infante_bridging_2010, Xu2020}, magnetic orders\cite{sando_crafting_2013, agbelele_strain_2016}, metal-insulator transitions\cite{Wang2019} and electro-optical properties\cite{BlazquezMartinez2023-Electro-optics, Paillard2019, Fredrickson2018} among others. 

In thin films, manipulating epitaxial strain is a well-known strategy to control the in-plane lattice parameters while the out-of-plane lattice parameters are imposed by the strain accommodation. The typical range of accessible in-plane strain yielded by epitaxy usually lies within 3-4\% of lattice mismatch\cite{Sando2022}. In epitaxial \bfo{} films, values of in-plane compressive strain larger than 4\% have been achieved by choosing the appropriate lattice mismatch between the film and the substrate to favor the transition towards the supertetragonal polymorph\cite{Zeches2009}. In polycrystalline films, which are more suited for functional applications due to their lower fabrication costs\cite{Muralt2000} and compatibility with low-temperature processing techniques\cite{Song2024}, the in-plane strain can also be controlled by the choice of substrate. This is typically achieved by either choosing substrates with a different thermal expansion coefficient \cite{Won2019}, or by deposition on flexible substrates\cite{YangBFMTO2019}. Reported strain values remain typically below 1\%\cite{Brennecka2004, Moreira2024, Schenk2021APL}. Since polycrystalline films are more brittle than their epitaxial counterparts, strain values larger than 0.5\% in perovskite compounds result in structural cracking,\cite{Coleman2019} making it challenging to engineer their functional properties through strain.

Negative pressure, to be understood as an increase of the volume of the material's unit cell\cite{Imre2007-NegativeP}, has been theoretically shown to be a very promising strategy to tune functional properties in materials\cite{liu_negative_2009, tinte_anomalous_2003, Aligia2001, sharma_designing_2017}. However, due to the technically challenging nature of achieving negative pressure, it has been experimentally under-explored. Chemical doping with substitution by species with higher ionic radius results in an increase in volume. Nevertheless, this modifies the chemical composition of the material and results in effects that are not purely due to the induced deformation\cite{Barazani2023}. Tensile in-plane strain, resulting from depositing thin films on substrates with higher lattice parameters, can also be seen as bi-axial negative strain\cite{Aguirre2004}. However, it can only induce discrete states of negative pressure, which depend on the chosen substrates and does not correspond to a continuously tunable parameter.

Previous studies have demonstrated that implanted helium can be used to introduce uniaxial out-of-plane negative pressure.\cite{guo_strain_2015, toulouse_patterning_2021} Being a noble gas, helium does not form bonds and implants interstitially, causing the material to expand. This volume increase can be tuned continuously with the implanted helium dose, up to the limit of amorphization. The amorphization threshold is found to be around $10^{16}$ He.cm$^{-2}$ for oxide perovskites\cite{livengood_subsurface_2009}.

\bfo{} is an extensively studied multiferroic material\cite{Heron2011}: it becomes ferroelectric below $T_C=1100$~K in the bulk\cite{lebeugle_very_2007}, and exhibits a G-type antiferromagnetic ordering of the Fe spins with a superimposed cycloidal modulation at temperatures below $T_N=640$~K \cite{lebeugle_room-temperature_2007}. In thin films, due to symmetry lowering, the usual rhombohedral $R3c$ structure of the bulk phase at ambient conditions becomes monoclinic, in a phase called R-like because of its similarity with the bulk rhombohedral phase. Under high compressive epitaxial strain, the \bfo{} thin films transition towards the so-called supertetragonal polymorph or T-like phase\cite{sando_multiferroic_2016} which is also in a monoclinic space group due to the symmetry lowering of the film geometry, and for which the ratio between the $c$ and $a$ lattice parameters reaches 1.23.

Under helium implantation, \bfo{} films transition towards this supertetragonal phase under lower epitaxial strain than virgin non-implanted films\cite{herklotz_designing_2019, toulouse_patterning_2021, Chen2019}. The nature of the transition (first or second order) is still debated, but an intermediate mixed phase with a coexistence of the R-like and the T-like phases is consistently observed. So far, only structural and no functional properties of \bfo{} films have been studied under negative pressure by helium implantation, except for a recent study showing the formation of antiphase domain walls in He-implanted \bfo{} films\cite{Cai2023}.

So far, all previous studies using helium implantation for strain engineering have been performed in epitaxial monocrystalline films. Until now, there have been no studies on the use of helium implantation for strain-engineering of polycrystalline films. Here, we demonstrate that we can induce a tunable uniaxial strain of up to 3.2\%. Remarkably, the ferroelectric properties of the ceramic remain stable at these high strain levels, underscoring the exceptional capability of helium implantation to achieve tunable strain in polycrystalline materials. This high strain value, coupled with the preservation of the microstructure and functional properties, is unprecedented in polycrystalline films; prior strategies only achieved strain tunability below 1\%, with cracks typically developing at strain levels higher than 0.5\%.\cite{Coleman2019, Brennecka2004, BlazquezMartinez2023, Moreira2024}

\section{Materials and methods}

\subsection{Thin films synthesis}

Polycrystalline and structurally textured \bfo{} films were grown by chemical solution deposition as described in a previous work\cite{BlazquezMartinez2021}. Films with different pristine strain states were studied. To do so, the films were deposited on substrates with different thermal expansion coefficients: $(100)$ MgO (Biotain Crystal, China), c-cut sapphire (Sap), and fused silica (FS) substrates (both Siegert Wafer, Germany). Before deposition, the substrates were coated with \SI{20}{nm} of non-ferroelectric atomic-layer-deposited \ch{HfO2} film to prevent diffusion during film fabrication, and a 13 nm-thick \ch{PbTiO3} seed layer to obtain $(100)_\mathrm{pc}$-textured films. The thickness of all the \bfo{} films was set to \SI[separate-uncertainty = true]{200(10)}{nm}. Both pure \bfo (BFO) and \bfo{} co-doped with 5\% manganese and 2\% titanium substituting on the Fe sites (BFMTO), more suited for electrical measurements due to their lower electrical leakage \cite{Qi2005GreatlyBiFeO3, Singh2007, Kawae2009ImprovedFilms, Raghavan2014EffectsFilms, BlazquezMartinez2021}, were studied in this work.

Four samples were used for the structural characterization, one pure \bfo{} film on fused silica (BFO/FS) as a reference and 5\% manganese and 2\% titanium co-doped \bfo{} films (BFMTO) on three different substrates: BFMTO/FS, BFMTO/Sap and BFMTO/MgO with different in-plane strain states reported previously as -0.22\% on MgO, +0.1\% on sapphire, and +0.38\% on fused silica\cite{BlazquezMartinez2023-Stress}. 

Interdigitated electrodes (IDEs) were used on BFMTO/Sap to measure the electrical properties in-plane\cite{BlazquezMartinezIDE2022, Aruchamy2022}: conventional lift-off photolithography and platinum sputtering were used to pattern the IDEs on top of the films. The IDEs were encapsulated using an epoxy-based photoresist (SU-8 3000, Kayacu Advanced Materials, USA) to prevent electric arcing between the fingers during the high voltage switching. To probe the out-of-plane electrical properties, we used a metal-insulator-metal geometry (MIM) in BFMTO/PtSi, where circular Pt top electrodes with \SI{100}{\um} diameter and \SI{100}{nm} thickness were patterned by lift-off photolithography and deposited by sputtering.

\begin{figure}[htpb]
\resizebox{8cm}{!}{\includegraphics{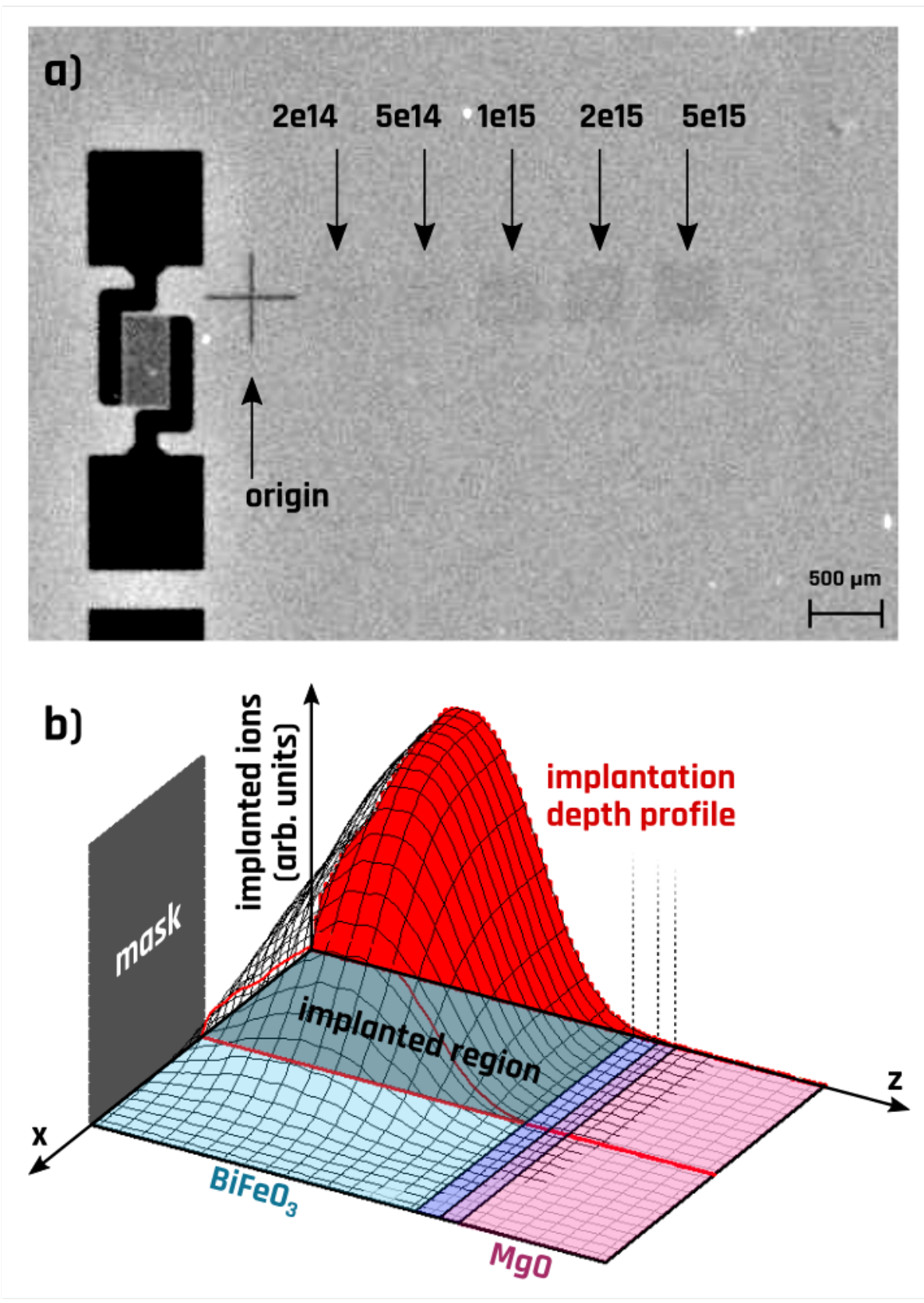}}
\caption{Helium implantation of the BFO/MgO film: (a) optical microscope image of the film after implantation. The implanted zones are visible with an increased optical contrast with the helium dose. (b) SRIM simulation of the implantation inside the \bfo{} film.}
\label{Fig1_implantation}
\end{figure}

\subsection{Helium implantation}

Local helium implantation was done using an Orion NanoFab Helium Ion Microscope (HIM) from Zeiss, which allows for a resolution for patterning down to the nanoscale\cite{zeiss_microscopy_2008}. For samples destined for synchrotron micro-XRD analysis, the patterning of the implanted regions was set to square regions of $450\times\SI{450}{\um\squared}$ to ensure that they remain much larger than the size of X-ray beam. For the samples destined for electrical measurements, we implanted regions corresponding to the surface needed for electrode deposition, namely $350\times\SI{750}{\um\squared}$ for the IDEs, and $200\times\SI{200}{\um\squared}$ for the MIM configuration.

Figure~\ref{Fig1_implantation}.a) shows an optical microscopy image of the BFO/MgO film after implantation. The different implanted regions are visible with a variable contrast. This change in the optical properties is likely due to surface pollution\cite{herklotz_optical_2019}, although the strong elasto-optic coupling in \bfo{} could also play a role in this optical change\cite{sando_linear_2014}. In any case, this change of contrast as well as the observed structural changes are still present twelve months after implantation.

The adequate implantation parameters, namely energy - that controls the implantation depth profile, shown in red on Figure~\ref{Fig1_implantation}.b - and the implanted dose - that controls the strain -, were determined using Monte Carlo simulations performed using the Stopping Range of Ions in Matter (SRIM) software package\cite{ziegler_srim-2003_2004}. The energy was set to \SI{21}{keV}, such that the maximum of the implantation depth profile is located in the center of the \bfo{} layer. The different samples were implanted with doses ranging from \SI{E14}{He.cm^{-2}} to \SI{5E15}{He.cm^{-2}}, chosen to be below the \SI{E16}{He.cm^{-2}} amorphization threshold for \bfo\cite{toulouse_patterning_2021}. The simulations were done with a mask to reproduce the implantation obtained with the scanning mode of the helium microscope and the in-plane dose gradient at the interface is visible along the $x$-axis. The size of the in-plane dose gradient, corresponding to the width of the interface between the implanted and non-implanted regions is around \SI{100}{\nm}.

\subsection{Structural and electrical characterization}

Micro-XRD measurements were performed on the CRISTAL beamline at the SOLEIL synchrotron source during two successive beamtimes. The wavelength of the beam was \SI{0.124512}{nm} and \SI{0.124743}{nm}, corresponding to an energy of \SI{9.96}{\keV} and \SI{9.94}{\keV} respectively.

Two different configurations were used to observe the effects of helium implantation: a reflection geometry, with the wavevector transfer near the normal of the film, to characterize the out-of-plane structure, and a transmission geometry where the wavevector transfer lies very close to the plane of the film. Due to the presence of some of the substrate's Bragg peaks close to the peaks of the layers we were observing (for instance the $(002)_{pc}$), we measured with a slight misalignment of 3$\si{\degree}$. \textcolor{black}{This may result in a change of relative intensity between different Bragg peaks but is otherwise inconsequential for the strain values, as explained in the supplementary information.}

Polarization hysteresis loops were measured on a Thin Film Analyzer TF 2000 (aixACCT, Germany) at room temperature and at a frequency of \SI{5}{kHz}. DC current density-electric field measurements were performed on unpoled films with a step of \SI{1.6}{\kV\per\cm} and a duration of \SI{2}{s} per step.

\section{Results and discussion} 

\subsection{Structural characterization}

Details about the microstructure and thermal strain analysis of the different films are identical to those reported previously\cite{BlazquezMartinez2023-Stress}. All films show a granular microstructure with a grain diameter around \SI{50}{nm}. The pristine in-plane residual thermal strain of the films grown on fused silica, sapphire, and MgO (0.38\%, 0.10\%, and -0.22\% respectively) results from the differences in thermal expansion coefficients.

\begin{figure*}[htpb]
\resizebox{\textwidth}{!}{\includegraphics{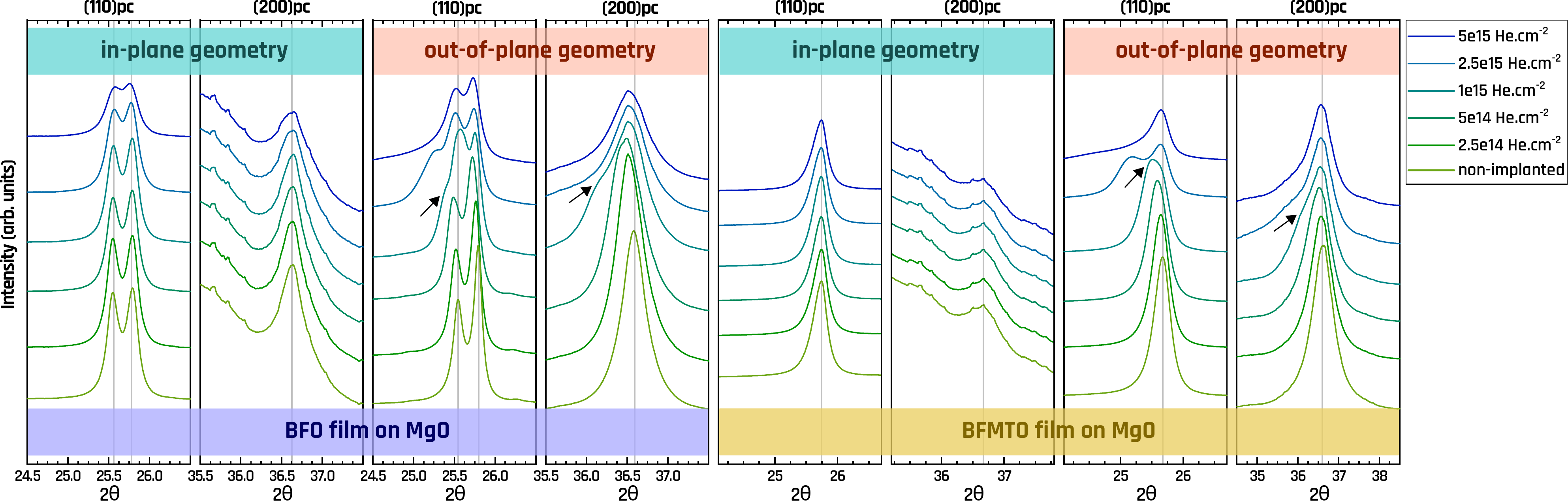}}
\caption{Some of the Bragg peaks measured in the pure and co-doped \bfo{} samples on MgO. The Bragg peaks measured on all samples can be found in Supplementary materials. The dotted vertical lines are guides for visualizing the peak shift. The arrows show the onset of phase mixing.}
\label{Fig2_XRD_tth}
\end{figure*}

To determine the effect of helium implantation on the structure of the \bfo{} films, micro-XRD measurements were performed on the different helium implanted regions on each sample. Figure~\ref{Fig2_XRD_tth} exemplarily shows the $\theta$--$2\theta$ scans of the $(110)_\mathrm{pc}$ and $(002)_\mathrm{pc}$ peak under different Helium doses for BFO/MgO and BFMTO/MgO. 

A shift towards lower \si{2\theta} angles, i.e. larger lattice parameter, with increasing implanted helium dose is observed. Under higher doses, the appearance of a new peak reveals the presence of a strained phase at lower \si{2\theta} angles with the persistence of the non-strained phase at the original \si{2\theta} resulting in phase-mixing, as expected for the structural transition towards the supertetragonal polymorph observed in epitaxial films under implantation\cite{herklotz_designing_2019, toulouse_patterning_2021}. The $\theta$--$2\theta$ scans for the different films with different doses are shown in the supplementary information.

\begin{figure*}[htpb]
\resizebox{\textwidth}{!}{\includegraphics{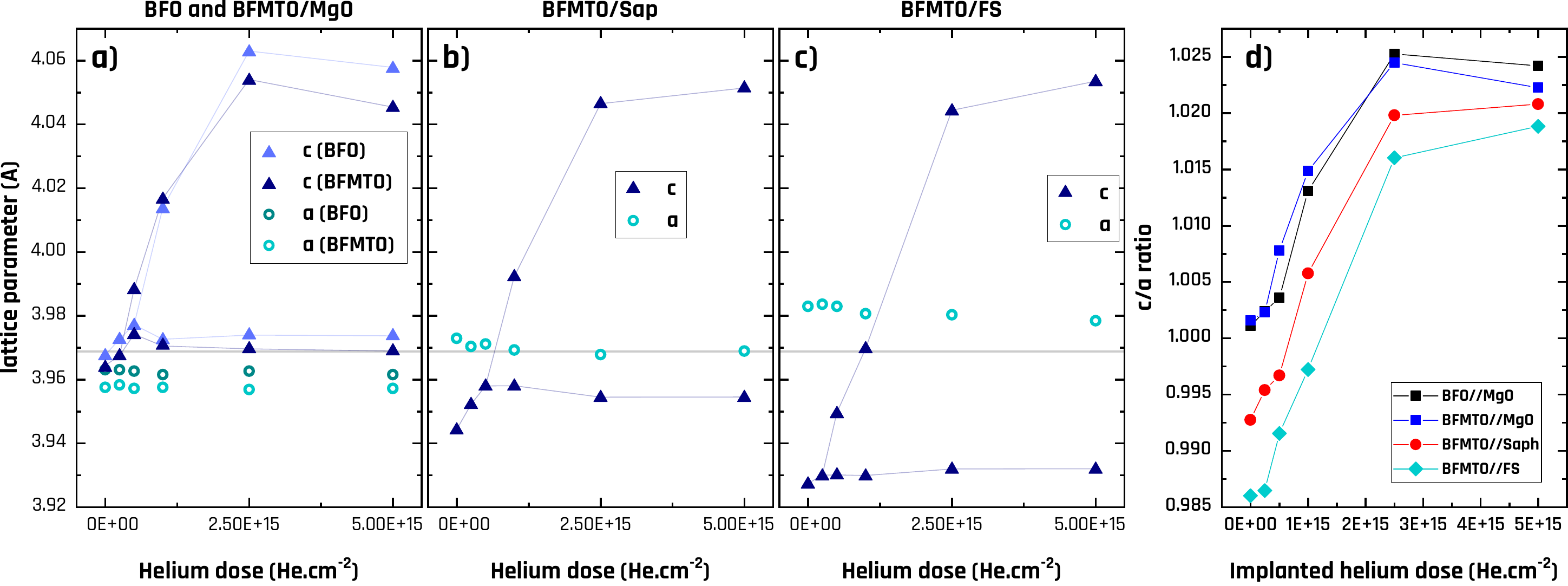}}
\caption{Lattice parameter evolution under helium implantation of (a) pure and co-doped \bfo{} films on MgO, and (b) and (c) co-doped \bfo{} films synthesized on sapphire and fused silica respectively. For higher helium doses, two values of the c-axis lattice parameter are given due to phase coexistence. The \bfo{} bulk reference of $a_{pc}$ is shown by the horizontal lines\cite{lebeugle_room-temperature_2007}. Error bars are smaller than the symbol size. (d) Evolution under helium implantation of the $c/a$ ratio of the strained phases in each sample.}
\label{Fig2_XRD_lp}
\end{figure*}

For each film, the in-plane (a) and out-of-plane (c) lattice parameters were extracted from the $(002)_\mathrm{pc}$ Bragg peak measured in reflection and transmission geometries. Figure~\ref{Fig2_XRD_lp} shows the lattice parameter evolution for different helium doses in films with different pristine strain states. The pristine strain state was evaluated in the non-implanted regions. Different $c/a$ ratios were measured in the different films as expected from the different thermal expansion coefficients of the substrates. The $c/a$ lattice parameter ratios are 1.003, 0.992 and 0.985 for BFMTO/MgO, BFMTO/Sap and BFMTO/FS, respectively. This is consistent with previously measured pristine strain states\cite{BlazquezMartinez2023-Stress}.

An out-of-plane expansion of the unit cell without any change in the plane is observed in all the samples with increasing helium dose. Slightly lower c-axis expansion with respect to the pristine state is observed in BFMTO compared to the BFO films. In all samples, the implantation-driven strain saturates around \SI{2.5E15}{He.cm^{-2}} dose. 

\textcolor{black}{In general, the width of the Bragg peaks tends to increase with the He dose, as can be seen in Fig.~\ref{Fig2_XRD_tth} and is further shown in representative plots of FWHM vs. dose (Figs.~S11-S14) in the supplementary information). Several factors can potentially contribute to peak broadening, most prominently the presence of strain gradients and a loss in crystallinity induced by implantation. Interestingly, the magnitude of this broadening varies strongly depending on the geometry. For measurements in the in-plane geometry, peak broadening remains very moderate (Fig.~S10). In co-doped sample where peaks are already broadened with respect to pure BFO, it is even hardly noticeable (Fig.~S11). This indicates that crystallinity is largely preserved. In the out-of-plane geometry, the Bragg peaks attributed to the unstrained fraction of the film also do not undergo a noticeable change in width but for the strained fraction of the films the broadening is much more significant (Figs.~S12 and S13). We attribute this to the presence of strain gradients in the thickness of the film deriving from the helium implantation depth profile.}


A clear influence of the pristine strain state is observed. Higher pristine in-plane strain results in higher out-of-plane lattice parameter increase with helium implantation. The measured c-axis expansion with saturation-level implantation is of 2.26\% on BFMTO/MgO (for 0.22\% compressive in-plane strain), 2.72\% on BFMTO/Sap (for 0.1\% tensile in-plane strain) and 3.21\% on BFMTO/FS (for 0.38\% tensile in-plane strain).

\subsection{Electrical properties}

The influence of He-implantation on the electrical properties was studied in two geometries: out-of-plane, using MIM capacitors of BFMTO on platinized silicon and in-plane, using IDE capacitors on BFMTO/Sap. 

\begin{figure*}[htpb]
\resizebox{\textwidth}{!}{\includegraphics{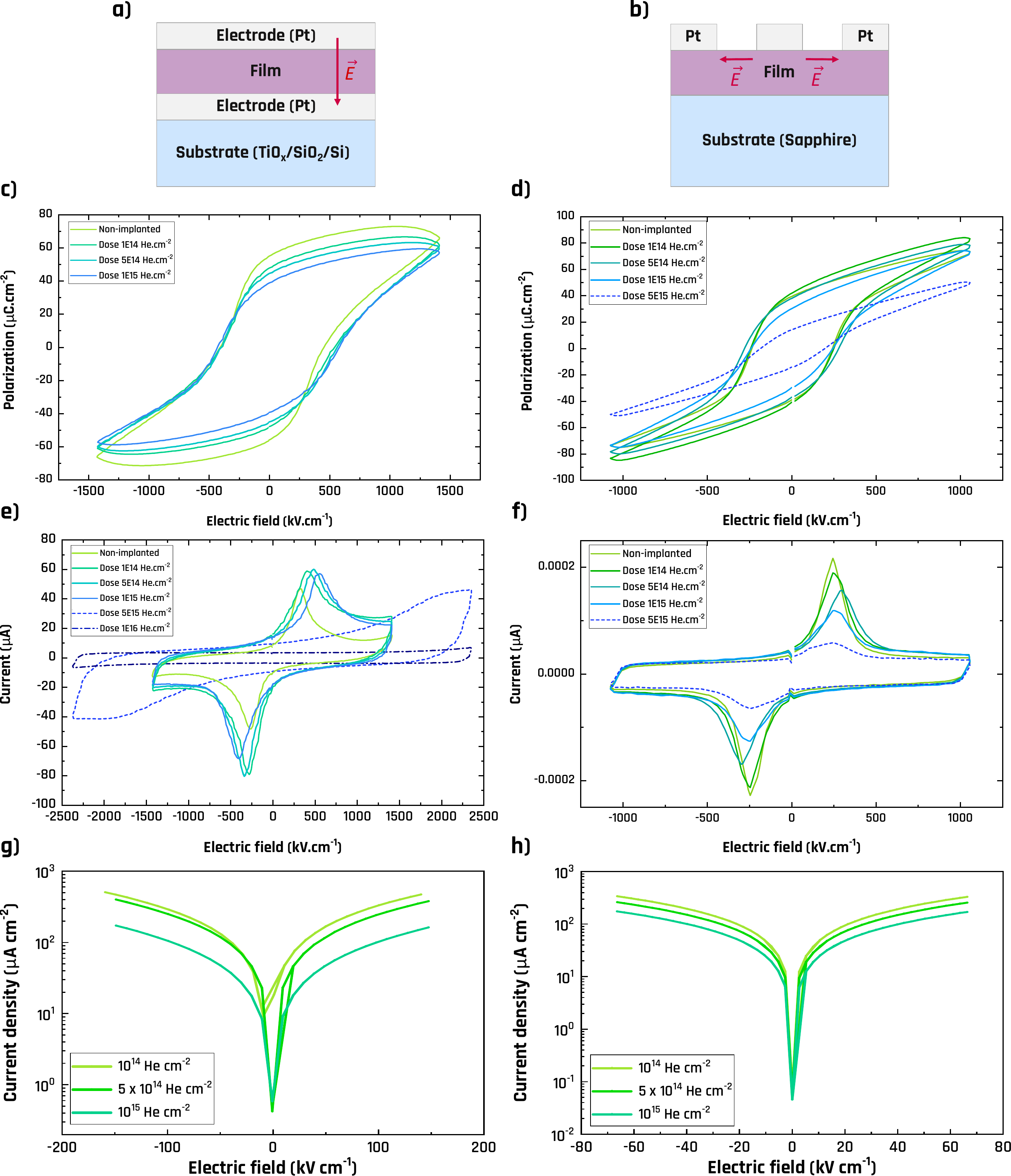}}
\caption{Device structure for the (a) metal-insulator-metal (MIM) configuration, applying the field out-of-plane, and (b) with interdigitated electrodes (IDE), applying the field in-plane. Hysteresis $P(E)$ loops for different helium doses taken in (c) MIM, and (d) IDE configurations and their associated $I(V)$ curves (e,f). Leakage current measurements for different helium doses taken in (g) MIM and (h) IDE configurations.}
\label{Fig3_PE}
\end{figure*}

The $P(E)$ curves of the Pt/BFMTO/Pt MIM capacitor as a function of the implanted He dose are displayed in Figure~\ref{Fig3_PE}.c. Stable ferroelectric switching is observed in all samples with doses up to \SI{E15}{He.cm^{-2}}. A small decrease of the remnant polarization P\textsubscript{r} is observed with increasing helium dose. At doses higher than \SI{5E15}{He.cm^{-2}}, the ferroelectric domain structure does not switch anymore (Fig.\ref{Fig3_PE}.e, dotted lines). This could be due to the progressive increase of the coercive electric field under helium (visible with the maxima of the $I(V)$ curves on Fig~\ref{Fig3_PE}.e). 

The in-plane switching characteristic using IDE geometry is shown in Figure~\ref{Fig3_PE}.d. A decrease in the remnant polarization P\textsubscript{r} is also observed with increasing helium dose. He-implantation doses higher than \SI{5E15}{He.cm^{-2}} also suppress the in-plane switching of the ferroelectric domain structure. However, unlike what is observed in the MIM geometry, the coercive switching field does not shift appreciably (Fig.~\ref{Fig3_PE}.f).

DC current density measurements under electric field  ($j$-$E$) were performed. To avoid contributions from displacement currents due to polarization switching, the curves were only measured at electric fields below 30\% of the coercive field. The $j$-$E$ curves for both configurations are displayed in Figure~\ref{Fig3_PE}.g and \ref{Fig3_PE}~.h. A reduction of the leakage current density is observed with helium implantation.

\subsection{Discussion}

Helium implantation allows for a continuously tunable increase of the out-of-plane lattice parameter in all the films. This indicates the transferability of the technique to strain engineer polycrystalline films, with a narrower strain range compared to their epitaxial counterparts\cite{guo_strain-induced_2018, toulouse_patterning_2021}. Despite a more complex microstructure and the presence of randomly oriented domains and grain boundaries, helium implantation still induces an out-of-plane strain with a lower out-of-plane lattice parameter expansion compared to epitaxial films. For epitaxial samples, an out-of-plane lattice parameter tunability up to 9\% has been reported in \bfo{} films deposited on \ch{SrTiO3}\cite{toulouse_patterning_2021}. In polycrystalline samples, the tunability only reaches a maximum value of 3.2\% in \bfo/FS. However, these values are one order of magnitude larger than what can be achieved using substrates with thermal expansion coefficient strongly dissimilar from that of the film, showing the potential of helium implantation as a strain engineering technique in polycrystalline samples. 
Contrary to what is observed in compressively strained epitaxial films, the onset of the transition towards the supertetragonal phase is not yet reached for the values of strain and $c/a$ ratio achieved here: up to 1.03 for the BFO/MgO film, well below the 1.23 of the T-phase. Our polycrystalline films stay in an R-like phase, with an out-of-plane elongation tunable under helium dose.

The pristine strain state plays a key role in the lattice parameter expansion upon helium implantation: the higher in-plane strain results in higher out-of-plane lattice parameter expansion. Additionally, the peak splitting in the x-ray diffractograms at higher doses indicates the coexistence of strained and non-strained regions in the film. In agreement with the SRIM simulations (Fig.~\ref{Fig1_implantation}.b), a skewed normal distribution of helium concentration is expected, resulting in a strain gradient within the film thickness that could explain this coexistence.

Stable ferroelectric switching is observed with helium doses up to \SI{E15}{He.cm^{-2}}. The absence of any strong change in the P\textsubscript{r} values is in agreement with the structure of the film remaining in a strained rhombohedral phase with no change in texture. In this phase, only subtle changes in the P\textsubscript{r} are expected \cite{daumont2012strain} and observed in this work, which supports the idea that helium affects strain only and does not lead to any significant modification of the dielectric - and here ferroelectric - properties in the material. \textcolor{black}{The increase in coercive field with increasing out-of-plane lattice parameter is consistent with previous findings for strained R-phase \bfo{} films. Phase field simulations\cite{biegalski2011strong} and experimental $P(E)$ loops\cite{daumont2012strain, jang2008strain} show that the coercive field strongly increases with increasing tetragonality. An increase exceeding 50\% has been observed for strain variations of 3\%, transitioning from tensile to compressive strain in films measured using metal-insulator-metal capacitors. In our case, the lower increase in the coercive field may arise from the nature of the strain-engineering method. In epitaxial strain both the out-of-plane and the in-plane lattice parameter change, and the volume remains constant. In our implanted films an uniaxial strain is induced, resulting in a volume expansion of the unit cell.}

\textcolor{black}{An additional mechanism that could induce an increase of coercive field is the formation of helium-implantation-induced defect complexes,\cite{Saremi2018} which are known to act as domain wall pinning centers, as reported for He-implanted BFO epitaxial films. However, the increase in coercive field observed in our samples is smaller than previously reported values for similar helium doses, indicating lower domain wall pinning with helium implantation.}

Additionally, a decrease of conductivity in the \bfo{} with increasing helium implantation is observed. This may not be a strain-induced effect, but could rather be attributed to the formation of defect complexes with increasing helium dose as previously shown in helium implanted epitaxial \ch{PbTiO3} films\cite{Saremi2016, saremi_local_2018}, or epitaxial \bfo\cite{Saremi2018} and epitaxial PMN-PT films\cite{Kim2020}.

\section{Conclusion and outlook}

We studied the influence of helium implantation on the structural and electrical properties of textured polycrystalline \bfo{} under different pristine strain states. An increase of the lattice parameter up to 3.2\% and 2.3\% in \bfo{} films under compressive and tensile strain respectively, was measured.

These results demonstrate a method to induce unprecedented large out-of-plane strain values in polycrystalline films, highlighting that grain boundaries do not prevent the interstitial implantation of helium atoms in the grains. The skewed normal distribution of helium concentration in the film induces differently strained areas through the thickness of the film, which results in the coexistence of strained and non-strained R-like regions. Because helium ion implantation can produce these high strain values in application-compatible polycrystalline thin films, it holds promise as a strategy to tune the strain-dependent properties of polycrystalline ceramics such as piezoelectricity or electro-optic coefficients. Additionally, the possibility to tune the helium concentration profile by modifying the implantation parameters such as energy or incident angle opens the path to inducing additional flexo-electric functionalities to polycrystalline materials.

\section*{Author contributions}
The samples were synthesized and electrodes were deposited by A.B.M. and S.G. SRIM simulations and determination of implantation parameters were done by C.T. Helium implantation was performed by J-N.A. and C.T. Electrical measurements were performed by A.B.M., T.G. and C.T. The synchrotron experiments on the CRISTAL beamline were done by A.B.M., M.G. and C.T. as users and P.F. as a beamline scientist and local contact. C.T. was the main proposer for both synchrotron proposals and performed the XRD data analysis. The paper was written by C.T., A.B.M and M.G, and was read and commented upon by all co-authors. C.T. coordinated the study.

\section*{Acknowledgments}
Supported by the Luxembourg National Research Fund (FNR) (FNR/C21/MS/16335086/Toulouse). A.B.M., S.G., and T.G. also acknowledge support by the Luxembourg National Research Fund (FNR) (PRIDE17/12246511/PACE).

We acknowledge the SOLEIL Synchrotron facility for the provision of synchrotron radiation at the CRISTAL beamline (proposals number 20210545 and 20221603).

This research was funded in whole, or in part, by the Luxembourg National Research Fund (FNR), grant reference FNR/C21/MS/16335086/Toulouse. For the purpose of open access, and in fulfilment of the obligations arising from the grant agreement, the author has applied a Creative Commons Attribution 4.0 International (CC BY 4.0) license to any Author Accepted Manuscript version arising from this submission.

The authors have no conflicts of interest to disclose.

The data that support the finding of this study are available from the corresponding author upon reasonable request.  

\newpage

\bibliography{BFO_sg_2024_paper}

\begin{thebibliography}{57}%
\makeatletter
\providecommand \@ifxundefined [1]{%
 \@ifx{#1\undefined}
}%
\providecommand \@ifnum [1]{%
 \ifnum #1\expandafter \@firstoftwo
 \else \expandafter \@secondoftwo
 \fi
}%
\providecommand \@ifx [1]{%
 \ifx #1\expandafter \@firstoftwo
 \else \expandafter \@secondoftwo
 \fi
}%
\providecommand \natexlab [1]{#1}%
\providecommand \enquote  [1]{``#1''}%
\providecommand \bibnamefont  [1]{#1}%
\providecommand \bibfnamefont [1]{#1}%
\providecommand \citenamefont [1]{#1}%
\providecommand \href@noop [0]{\@secondoftwo}%
\providecommand \href [0]{\begingroup \@sanitize@url \@href}%
\providecommand \@href[1]{\@@startlink{#1}\@@href}%
\providecommand \@@href[1]{\endgroup#1\@@endlink}%
\providecommand \@sanitize@url [0]{\catcode `\\12\catcode `\$12\catcode `\&12\catcode `\#12\catcode `\^12\catcode `\_12\catcode `\%12\relax}%
\providecommand \@@startlink[1]{}%
\providecommand \@@endlink[0]{}%
\providecommand \url  [0]{\begingroup\@sanitize@url \@url }%
\providecommand \@url [1]{\endgroup\@href {#1}{\urlprefix }}%
\providecommand \urlprefix  [0]{URL }%
\providecommand \Eprint [0]{\href }%
\providecommand \doibase [0]{http://dx.doi.org/}%
\providecommand \selectlanguage [0]{\@gobble}%
\providecommand \bibinfo  [0]{\@secondoftwo}%
\providecommand \bibfield  [0]{\@secondoftwo}%
\providecommand \translation [1]{[#1]}%
\providecommand \BibitemOpen [0]{}%
\providecommand \bibitemStop [0]{}%
\providecommand \bibitemNoStop [0]{.\EOS\space}%
\providecommand \EOS [0]{\spacefactor3000\relax}%
\providecommand \BibitemShut  [1]{\csname bibitem#1\endcsname}%
\let\auto@bib@innerbib\@empty
\bibitem [{\citenamefont {Choi}\ \emph {et~al.}(2004)\citenamefont {Choi}, \citenamefont {Biegalski}, \citenamefont {Li}, \citenamefont {Sharan}, \citenamefont {Schubert}, \citenamefont {Uecker}, \citenamefont {Reiche}, \citenamefont {Chen}, \citenamefont {Pan}, \citenamefont {Gopalan}, \citenamefont {Chen}, \citenamefont {Schlom},\ and\ \citenamefont {Eom}}]{Choi2004}%
  \BibitemOpen
  \bibfield  {author} {\bibinfo {author} {\bibfnamefont {K.~J.}\ \bibnamefont {Choi}}, \bibinfo {author} {\bibfnamefont {M.}~\bibnamefont {Biegalski}}, \bibinfo {author} {\bibfnamefont {Y.~L.}\ \bibnamefont {Li}}, \bibinfo {author} {\bibfnamefont {A.}~\bibnamefont {Sharan}}, \bibinfo {author} {\bibfnamefont {J.}~\bibnamefont {Schubert}}, \bibinfo {author} {\bibfnamefont {R.}~\bibnamefont {Uecker}}, \bibinfo {author} {\bibfnamefont {P.}~\bibnamefont {Reiche}}, \bibinfo {author} {\bibfnamefont {Y.~B.}\ \bibnamefont {Chen}}, \bibinfo {author} {\bibfnamefont {X.~Q.}\ \bibnamefont {Pan}}, \bibinfo {author} {\bibfnamefont {V.}~\bibnamefont {Gopalan}}, \bibinfo {author} {\bibfnamefont {L.-Q.}\ \bibnamefont {Chen}}, \bibinfo {author} {\bibfnamefont {D.~G.}\ \bibnamefont {Schlom}}, \ and\ \bibinfo {author} {\bibfnamefont {C.~B.}\ \bibnamefont {Eom}},\ }\bibfield  {title} {\enquote {\bibinfo {title} {Enhancement of ferroelectricity in strained {BaTiO$_3$} thin films},}\ }\href {\doibase 10.1126/science.1103218}
  {\bibfield  {journal} {\bibinfo  {journal} {Science}\ }\textbf {\bibinfo {volume} {306}},\ \bibinfo {pages} {1005--1009} (\bibinfo {year} {2004})}\BibitemShut {NoStop}%
\bibitem [{\citenamefont {Infante}\ \emph {et~al.}(2010)\citenamefont {Infante}, \citenamefont {Lisenkov}, \citenamefont {Dupé}, \citenamefont {Bibes}, \citenamefont {Fusil}, \citenamefont {Jacquet}, \citenamefont {Geneste}, \citenamefont {Petit}, \citenamefont {Courtial}, \citenamefont {Juraszek}, \citenamefont {Bellaiche}, \citenamefont {Barthélémy},\ and\ \citenamefont {Dkhil}}]{infante_bridging_2010}%
  \BibitemOpen
  \bibfield  {author} {\bibinfo {author} {\bibfnamefont {I.~C.}\ \bibnamefont {Infante}}, \bibinfo {author} {\bibfnamefont {S.}~\bibnamefont {Lisenkov}}, \bibinfo {author} {\bibfnamefont {B.}~\bibnamefont {Dupé}}, \bibinfo {author} {\bibfnamefont {M.}~\bibnamefont {Bibes}}, \bibinfo {author} {\bibfnamefont {S.}~\bibnamefont {Fusil}}, \bibinfo {author} {\bibfnamefont {E.}~\bibnamefont {Jacquet}}, \bibinfo {author} {\bibfnamefont {G.}~\bibnamefont {Geneste}}, \bibinfo {author} {\bibfnamefont {S.}~\bibnamefont {Petit}}, \bibinfo {author} {\bibfnamefont {A.}~\bibnamefont {Courtial}}, \bibinfo {author} {\bibfnamefont {J.}~\bibnamefont {Juraszek}}, \bibinfo {author} {\bibfnamefont {L.}~\bibnamefont {Bellaiche}}, \bibinfo {author} {\bibfnamefont {A.}~\bibnamefont {Barthélémy}}, \ and\ \bibinfo {author} {\bibfnamefont {B.}~\bibnamefont {Dkhil}},\ }\bibfield  {title} {\enquote {\bibinfo {title} {Bridging {Multiferroic} {Phase} {Transitions} by {Epitaxial} {Strain} in {BiFeO$_3$}},}\ }\href {\doibase
  10.1103/PhysRevLett.105.057601} {\bibfield  {journal} {\bibinfo  {journal} {Physical Review Letters}\ }\textbf {\bibinfo {volume} {105}} (\bibinfo {year} {2010}),\ 10.1103/PhysRevLett.105.057601}\BibitemShut {NoStop}%
\bibitem [{\citenamefont {Xu}\ \emph {et~al.}(2020)\citenamefont {Xu}, \citenamefont {Huang}, \citenamefont {Barnard}, \citenamefont {Hong}, \citenamefont {Singh}, \citenamefont {Wong}, \citenamefont {Jansen}, \citenamefont {Harbola}, \citenamefont {Xiao}, \citenamefont {Wang}, \citenamefont {Crossley}, \citenamefont {Lu}, \citenamefont {Liu},\ and\ \citenamefont {Hwang}}]{Xu2020}%
  \BibitemOpen
  \bibfield  {author} {\bibinfo {author} {\bibfnamefont {R.}~\bibnamefont {Xu}}, \bibinfo {author} {\bibfnamefont {J.}~\bibnamefont {Huang}}, \bibinfo {author} {\bibfnamefont {E.~S.}\ \bibnamefont {Barnard}}, \bibinfo {author} {\bibfnamefont {S.~S.}\ \bibnamefont {Hong}}, \bibinfo {author} {\bibfnamefont {P.}~\bibnamefont {Singh}}, \bibinfo {author} {\bibfnamefont {E.~K.}\ \bibnamefont {Wong}}, \bibinfo {author} {\bibfnamefont {T.}~\bibnamefont {Jansen}}, \bibinfo {author} {\bibfnamefont {V.}~\bibnamefont {Harbola}}, \bibinfo {author} {\bibfnamefont {J.}~\bibnamefont {Xiao}}, \bibinfo {author} {\bibfnamefont {B.~Y.}\ \bibnamefont {Wang}}, \bibinfo {author} {\bibfnamefont {S.}~\bibnamefont {Crossley}}, \bibinfo {author} {\bibfnamefont {D.}~\bibnamefont {Lu}}, \bibinfo {author} {\bibfnamefont {S.}~\bibnamefont {Liu}}, \ and\ \bibinfo {author} {\bibfnamefont {H.~Y.}\ \bibnamefont {Hwang}},\ }\bibfield  {title} {\enquote {\bibinfo {title} {Strain-induced room-temperature ferroelectricity in {SrTiO$_3$}
  membranes},}\ }\href {\doibase 10.1038/s41467-020-16912-3} {\bibfield  {journal} {\bibinfo  {journal} {Nature Communications}\ }\textbf {\bibinfo {volume} {11}},\ \bibinfo {pages} {3141} (\bibinfo {year} {2020})}\BibitemShut {NoStop}%
\bibitem [{\citenamefont {Sando}\ \emph {et~al.}(2013)\citenamefont {Sando}, \citenamefont {Agbelele}, \citenamefont {Rahmedov}, \citenamefont {Liu}, \citenamefont {Rovillain}, \citenamefont {Toulouse}, \citenamefont {Infante}, \citenamefont {Pyatakov}, \citenamefont {Fusil}, \citenamefont {Jacquet}, \citenamefont {Carrétéro}, \citenamefont {Deranlot}, \citenamefont {Lisenkov}, \citenamefont {Wang}, \citenamefont {Le~Breton}, \citenamefont {Cazayous}, \citenamefont {Sacuto}, \citenamefont {Juraszek}, \citenamefont {Zvezdin}, \citenamefont {Bellaiche}, \citenamefont {Dkhil}, \citenamefont {Barthélémy},\ and\ \citenamefont {Bibes}}]{sando_crafting_2013}%
  \BibitemOpen
  \bibfield  {author} {\bibinfo {author} {\bibfnamefont {D.}~\bibnamefont {Sando}}, \bibinfo {author} {\bibfnamefont {A.}~\bibnamefont {Agbelele}}, \bibinfo {author} {\bibfnamefont {D.}~\bibnamefont {Rahmedov}}, \bibinfo {author} {\bibfnamefont {J.}~\bibnamefont {Liu}}, \bibinfo {author} {\bibfnamefont {P.}~\bibnamefont {Rovillain}}, \bibinfo {author} {\bibfnamefont {C.}~\bibnamefont {Toulouse}}, \bibinfo {author} {\bibfnamefont {I.~C.}\ \bibnamefont {Infante}}, \bibinfo {author} {\bibfnamefont {A.~P.}\ \bibnamefont {Pyatakov}}, \bibinfo {author} {\bibfnamefont {S.}~\bibnamefont {Fusil}}, \bibinfo {author} {\bibfnamefont {E.}~\bibnamefont {Jacquet}}, \bibinfo {author} {\bibfnamefont {C.}~\bibnamefont {Carrétéro}}, \bibinfo {author} {\bibfnamefont {C.}~\bibnamefont {Deranlot}}, \bibinfo {author} {\bibfnamefont {S.}~\bibnamefont {Lisenkov}}, \bibinfo {author} {\bibfnamefont {D.}~\bibnamefont {Wang}}, \bibinfo {author} {\bibfnamefont {J.-M.}\ \bibnamefont {Le~Breton}}, \bibinfo {author} {\bibfnamefont
  {M.}~\bibnamefont {Cazayous}}, \bibinfo {author} {\bibfnamefont {A.}~\bibnamefont {Sacuto}}, \bibinfo {author} {\bibfnamefont {J.}~\bibnamefont {Juraszek}}, \bibinfo {author} {\bibfnamefont {A.~K.}\ \bibnamefont {Zvezdin}}, \bibinfo {author} {\bibfnamefont {L.}~\bibnamefont {Bellaiche}}, \bibinfo {author} {\bibfnamefont {B.}~\bibnamefont {Dkhil}}, \bibinfo {author} {\bibfnamefont {A.}~\bibnamefont {Barthélémy}}, \ and\ \bibinfo {author} {\bibfnamefont {M.}~\bibnamefont {Bibes}},\ }\bibfield  {title} {{\selectlanguage {en}\enquote {\bibinfo {title} {Crafting the magnonic and spintronic response of bifeo$_3$ films by epitaxial strain},}\ }}\href {\doibase 10.1038/nmat3629} {\bibfield  {journal} {\bibinfo  {journal} {Nature Materials}\ }\textbf {\bibinfo {volume} {12}},\ \bibinfo {pages} {641--646} (\bibinfo {year} {2013})}\BibitemShut {NoStop}%
\bibitem [{\citenamefont {Agbelele}\ \emph {et~al.}(2016)\citenamefont {Agbelele}, \citenamefont {Sando}, \citenamefont {Toulouse}, \citenamefont {Paillard}, \citenamefont {Johnson}, \citenamefont {Rüffer}, \citenamefont {Popkov}, \citenamefont {Carrétéro}, \citenamefont {Rovillain}, \citenamefont {Le~Breton}, \citenamefont {Dkhil}, \citenamefont {Cazayous}, \citenamefont {Gallais}, \citenamefont {Méasson}, \citenamefont {Sacuto}, \citenamefont {Manuel}, \citenamefont {Zvezdin}, \citenamefont {Barthélémy}, \citenamefont {Juraszek},\ and\ \citenamefont {Bibes}}]{agbelele_strain_2016}%
  \BibitemOpen
  \bibfield  {author} {\bibinfo {author} {\bibfnamefont {A.}~\bibnamefont {Agbelele}}, \bibinfo {author} {\bibfnamefont {D.}~\bibnamefont {Sando}}, \bibinfo {author} {\bibfnamefont {C.}~\bibnamefont {Toulouse}}, \bibinfo {author} {\bibfnamefont {C.}~\bibnamefont {Paillard}}, \bibinfo {author} {\bibfnamefont {R.~D.}\ \bibnamefont {Johnson}}, \bibinfo {author} {\bibfnamefont {R.}~\bibnamefont {Rüffer}}, \bibinfo {author} {\bibfnamefont {A.~F.}\ \bibnamefont {Popkov}}, \bibinfo {author} {\bibfnamefont {C.}~\bibnamefont {Carrétéro}}, \bibinfo {author} {\bibfnamefont {P.}~\bibnamefont {Rovillain}}, \bibinfo {author} {\bibfnamefont {J.-M.}\ \bibnamefont {Le~Breton}}, \bibinfo {author} {\bibfnamefont {B.}~\bibnamefont {Dkhil}}, \bibinfo {author} {\bibfnamefont {M.}~\bibnamefont {Cazayous}}, \bibinfo {author} {\bibfnamefont {Y.}~\bibnamefont {Gallais}}, \bibinfo {author} {\bibfnamefont {M.-A.}\ \bibnamefont {Méasson}}, \bibinfo {author} {\bibfnamefont {A.}~\bibnamefont {Sacuto}}, \bibinfo {author} {\bibfnamefont
  {P.}~\bibnamefont {Manuel}}, \bibinfo {author} {\bibfnamefont {A.~K.}\ \bibnamefont {Zvezdin}}, \bibinfo {author} {\bibfnamefont {A.}~\bibnamefont {Barthélémy}}, \bibinfo {author} {\bibfnamefont {J.}~\bibnamefont {Juraszek}}, \ and\ \bibinfo {author} {\bibfnamefont {M.}~\bibnamefont {Bibes}},\ }\bibfield  {title} {{\selectlanguage {en}\enquote {\bibinfo {title} {Strain and {Magnetic} {Field} {Induced} {Spin}-{Structure} {Transitions} in {Multiferroic} {BiFeO$_3$}},}\ }}\href {\doibase 10.1002/adma.201602327} {\bibfield  {journal} {\bibinfo  {journal} {Advanced Materials}\ }\textbf {\bibinfo {volume} {29}},\ \bibinfo {pages} {1602327} (\bibinfo {year} {2016})}\BibitemShut {NoStop}%
\bibitem [{\citenamefont {Wang}\ \emph {et~al.}(2019)\citenamefont {Wang}, \citenamefont {Zhang}, \citenamefont {Deepak}, \citenamefont {Chen}, \citenamefont {Fouchet}, \citenamefont {Duan}, \citenamefont {Hilliard}, \citenamefont {Kentsch}, \citenamefont {Chen}, \citenamefont {Zeng}, \citenamefont {Gao}, \citenamefont {Zeng}, \citenamefont {Helm}, \citenamefont {Prellier},\ and\ \citenamefont {Zhou}}]{Wang2019}%
  \BibitemOpen
  \bibfield  {author} {\bibinfo {author} {\bibfnamefont {C.}~\bibnamefont {Wang}}, \bibinfo {author} {\bibfnamefont {H.}~\bibnamefont {Zhang}}, \bibinfo {author} {\bibfnamefont {K.}~\bibnamefont {Deepak}}, \bibinfo {author} {\bibfnamefont {C.}~\bibnamefont {Chen}}, \bibinfo {author} {\bibfnamefont {A.}~\bibnamefont {Fouchet}}, \bibinfo {author} {\bibfnamefont {J.}~\bibnamefont {Duan}}, \bibinfo {author} {\bibfnamefont {D.}~\bibnamefont {Hilliard}}, \bibinfo {author} {\bibfnamefont {U.}~\bibnamefont {Kentsch}}, \bibinfo {author} {\bibfnamefont {D.}~\bibnamefont {Chen}}, \bibinfo {author} {\bibfnamefont {M.}~\bibnamefont {Zeng}}, \bibinfo {author} {\bibfnamefont {X.}~\bibnamefont {Gao}}, \bibinfo {author} {\bibfnamefont {Y.-J.}\ \bibnamefont {Zeng}}, \bibinfo {author} {\bibfnamefont {M.}~\bibnamefont {Helm}}, \bibinfo {author} {\bibfnamefont {W.}~\bibnamefont {Prellier}}, \ and\ \bibinfo {author} {\bibfnamefont {S.}~\bibnamefont {Zhou}},\ }\bibfield  {title} {\enquote {\bibinfo {title} {Tuning the
  metal-insulator transition in epitaxial {SrVO$_3$} films by uniaxial strain},}\ }\href {\doibase 10.1103/PhysRevMaterials.3.115001} {\bibfield  {journal} {\bibinfo  {journal} {Physical Review Materials}\ }\textbf {\bibinfo {volume} {3}},\ \bibinfo {pages} {115001} (\bibinfo {year} {2019})}\BibitemShut {NoStop}%
\bibitem [{\citenamefont {Bl\'azquez~Mart\'inez}\ \emph {et~al.}(2023{\natexlab{a}})\citenamefont {Bl\'azquez~Mart\'inez}, \citenamefont {Grysan}, \citenamefont {Girod}, \citenamefont {Glin\v{s}ek}, \citenamefont {Aruchamy}, \citenamefont {Biswas}, \citenamefont {Guennou},\ and\ \citenamefont {Granzow}}]{BlazquezMartinez2023-Electro-optics}%
  \BibitemOpen
  \bibfield  {author} {\bibinfo {author} {\bibfnamefont {A.}~\bibnamefont {Bl\'azquez~Mart\'inez}}, \bibinfo {author} {\bibfnamefont {P.}~\bibnamefont {Grysan}}, \bibinfo {author} {\bibfnamefont {S.}~\bibnamefont {Girod}}, \bibinfo {author} {\bibfnamefont {S.}~\bibnamefont {Glin\v{s}ek}}, \bibinfo {author} {\bibfnamefont {N.}~\bibnamefont {Aruchamy}}, \bibinfo {author} {\bibfnamefont {P.}~\bibnamefont {Biswas}}, \bibinfo {author} {\bibfnamefont {M.}~\bibnamefont {Guennou}}, \ and\ \bibinfo {author} {\bibfnamefont {T.}~\bibnamefont {Granzow}},\ }\bibfield  {title} {\enquote {\bibinfo {title} {Strain engineering of the electro-optic effect in polycrystalline {BiFeO$_3$} films},}\ }\href {\doibase 10.1364/OME.485892} {\bibfield  {journal} {\bibinfo  {journal} {Optical Materials Express}\ }\textbf {\bibinfo {volume} {13}},\ \bibinfo {pages} {2061} (\bibinfo {year} {2023}{\natexlab{a}})}\BibitemShut {NoStop}%
\bibitem [{\citenamefont {Paillard}, \citenamefont {Prokhorenko},\ and\ \citenamefont {Bellaiche}(2019)}]{Paillard2019}%
  \BibitemOpen
  \bibfield  {author} {\bibinfo {author} {\bibfnamefont {C.}~\bibnamefont {Paillard}}, \bibinfo {author} {\bibfnamefont {S.}~\bibnamefont {Prokhorenko}}, \ and\ \bibinfo {author} {\bibfnamefont {L.}~\bibnamefont {Bellaiche}},\ }\bibfield  {title} {\enquote {\bibinfo {title} {Strain engineering of electro-optic constants in ferroelectric materials},}\ }\href {\doibase 10.1038/s41524-018-0141-4} {\bibfield  {journal} {\bibinfo  {journal} {npj Computational Materials}\ }\textbf {\bibinfo {volume} {5}},\ \bibinfo {pages} {6} (\bibinfo {year} {2019})}\BibitemShut {NoStop}%
\bibitem [{\citenamefont {Fredrickson}\ \emph {et~al.}(2018)\citenamefont {Fredrickson}, \citenamefont {Vogler-Neuling}, \citenamefont {Kormondy}, \citenamefont {Caimi}, \citenamefont {Eltes}, \citenamefont {Sousa}, \citenamefont {Fompeyrine}, \citenamefont {Abel},\ and\ \citenamefont {Demkov}}]{Fredrickson2018}%
  \BibitemOpen
  \bibfield  {author} {\bibinfo {author} {\bibfnamefont {K.~D.}\ \bibnamefont {Fredrickson}}, \bibinfo {author} {\bibfnamefont {V.~V.}\ \bibnamefont {Vogler-Neuling}}, \bibinfo {author} {\bibfnamefont {K.~J.}\ \bibnamefont {Kormondy}}, \bibinfo {author} {\bibfnamefont {D.}~\bibnamefont {Caimi}}, \bibinfo {author} {\bibfnamefont {F.}~\bibnamefont {Eltes}}, \bibinfo {author} {\bibfnamefont {M.}~\bibnamefont {Sousa}}, \bibinfo {author} {\bibfnamefont {J.}~\bibnamefont {Fompeyrine}}, \bibinfo {author} {\bibfnamefont {S.}~\bibnamefont {Abel}}, \ and\ \bibinfo {author} {\bibfnamefont {A.~A.}\ \bibnamefont {Demkov}},\ }\bibfield  {title} {\enquote {\bibinfo {title} {Strain enhancement of the electro-optical response in {BaTiO$_3$} films integrated on {S}i(001)},}\ }\href {\doibase 10.1103/PhysRevB.98.075136} {\bibfield  {journal} {\bibinfo  {journal} {Physical Review B}\ }\textbf {\bibinfo {volume} {98}},\ \bibinfo {pages} {075136} (\bibinfo {year} {2018})}\BibitemShut {NoStop}%
\bibitem [{\citenamefont {Sando}(2022)}]{Sando2022}%
  \BibitemOpen
  \bibfield  {author} {\bibinfo {author} {\bibfnamefont {D.}~\bibnamefont {Sando}},\ }\bibfield  {title} {\enquote {\bibinfo {title} {Strain and orientation engineering in {ABO$_3$} perovskite oxide thin films},}\ }\href {\doibase 10.1088/1361-648X/ac4c61} {\bibfield  {journal} {\bibinfo  {journal} {Journal of Physics: Condensed Matter}\ }\textbf {\bibinfo {volume} {34}},\ \bibinfo {pages} {153001} (\bibinfo {year} {2022})}\BibitemShut {NoStop}%
\bibitem [{\citenamefont {Zeches}\ \emph {et~al.}(2009)\citenamefont {Zeches}, \citenamefont {Rossell}, \citenamefont {Zhang}, \citenamefont {Hatt}, \citenamefont {He}, \citenamefont {Yang}, \citenamefont {Kumar}, \citenamefont {Wang}, \citenamefont {Melville}, \citenamefont {Adamo}, \citenamefont {Sheng}, \citenamefont {Chu}, \citenamefont {Ihlefeld}, \citenamefont {Erni}, \citenamefont {Ederer}, \citenamefont {Gopalan}, \citenamefont {Chen}, \citenamefont {Schlom}, \citenamefont {Spaldin}, \citenamefont {Martin},\ and\ \citenamefont {Ramesh}}]{Zeches2009}%
  \BibitemOpen
  \bibfield  {author} {\bibinfo {author} {\bibfnamefont {R.~J.}\ \bibnamefont {Zeches}}, \bibinfo {author} {\bibfnamefont {M.~D.}\ \bibnamefont {Rossell}}, \bibinfo {author} {\bibfnamefont {J.~X.}\ \bibnamefont {Zhang}}, \bibinfo {author} {\bibfnamefont {A.~J.}\ \bibnamefont {Hatt}}, \bibinfo {author} {\bibfnamefont {Q.}~\bibnamefont {He}}, \bibinfo {author} {\bibfnamefont {C.-H.}\ \bibnamefont {Yang}}, \bibinfo {author} {\bibfnamefont {A.}~\bibnamefont {Kumar}}, \bibinfo {author} {\bibfnamefont {C.~H.}\ \bibnamefont {Wang}}, \bibinfo {author} {\bibfnamefont {A.}~\bibnamefont {Melville}}, \bibinfo {author} {\bibfnamefont {C.}~\bibnamefont {Adamo}}, \bibinfo {author} {\bibfnamefont {G.}~\bibnamefont {Sheng}}, \bibinfo {author} {\bibfnamefont {Y.-H.}\ \bibnamefont {Chu}}, \bibinfo {author} {\bibfnamefont {J.~F.}\ \bibnamefont {Ihlefeld}}, \bibinfo {author} {\bibfnamefont {R.}~\bibnamefont {Erni}}, \bibinfo {author} {\bibfnamefont {C.}~\bibnamefont {Ederer}}, \bibinfo {author} {\bibfnamefont {V.}~\bibnamefont
  {Gopalan}}, \bibinfo {author} {\bibfnamefont {L.~Q.}\ \bibnamefont {Chen}}, \bibinfo {author} {\bibfnamefont {D.~G.}\ \bibnamefont {Schlom}}, \bibinfo {author} {\bibfnamefont {N.~A.}\ \bibnamefont {Spaldin}}, \bibinfo {author} {\bibfnamefont {L.~W.}\ \bibnamefont {Martin}}, \ and\ \bibinfo {author} {\bibfnamefont {R.}~\bibnamefont {Ramesh}},\ }\bibfield  {title} {\enquote {\bibinfo {title} {A strain-driven morphotropic phase boundary in {BiFeO$_3$}},}\ }\href {\doibase 10.1126/science.1177046} {\bibfield  {journal} {\bibinfo  {journal} {Science}\ }\textbf {\bibinfo {volume} {326}},\ \bibinfo {pages} {977--980} (\bibinfo {year} {2009})}\BibitemShut {NoStop}%
\bibitem [{\citenamefont {Muralt}(2000)}]{Muralt2000}%
  \BibitemOpen
  \bibfield  {author} {\bibinfo {author} {\bibfnamefont {P.}~\bibnamefont {Muralt}},\ }\bibfield  {title} {\enquote {\bibinfo {title} {Ferroelectric thin films for micro-sensors and actuators: a review},}\ }\href {\doibase 10.1088/0960-1317/10/2/307} {\bibfield  {journal} {\bibinfo  {journal} {Journal of Micromechanics and Microengineering}\ }\textbf {\bibinfo {volume} {10}},\ \bibinfo {pages} {136--146} (\bibinfo {year} {2000})}\BibitemShut {NoStop}%
\bibitem [{\citenamefont {Song}\ \emph {et~al.}(2024)\citenamefont {Song}, \citenamefont {Cardoletti}, \citenamefont {Mart\'inez}, \citenamefont {Ben\v{c}an}, \citenamefont {Kmet}, \citenamefont {Girod}, \citenamefont {Defay},\ and\ \citenamefont {Glin\v{s}ek}}]{Song2024}%
  \BibitemOpen
  \bibfield  {author} {\bibinfo {author} {\bibfnamefont {L.}~\bibnamefont {Song}}, \bibinfo {author} {\bibfnamefont {J.}~\bibnamefont {Cardoletti}}, \bibinfo {author} {\bibfnamefont {A.~B.}\ \bibnamefont {Mart\'inez}}, \bibinfo {author} {\bibfnamefont {A.}~\bibnamefont {Ben\v{c}an}}, \bibinfo {author} {\bibfnamefont {B.}~\bibnamefont {Kmet}}, \bibinfo {author} {\bibfnamefont {S.}~\bibnamefont {Girod}}, \bibinfo {author} {\bibfnamefont {E.}~\bibnamefont {Defay}}, \ and\ \bibinfo {author} {\bibfnamefont {S.}~\bibnamefont {Glin\v{s}ek}},\ }\bibfield  {title} {\enquote {\bibinfo {title} {Crystallization of piezoceramic films on glass via flash lamp annealing},}\ }\href {\doibase 10.1038/s41467-024-46257-0} {\bibfield  {journal} {\bibinfo  {journal} {Nature Communications}\ }\textbf {\bibinfo {volume} {15}},\ \bibinfo {pages} {1890} (\bibinfo {year} {2024})}\BibitemShut {NoStop}%
\bibitem [{\citenamefont {Won}\ \emph {et~al.}(2019)\citenamefont {Won}, \citenamefont {Seo}, \citenamefont {Kawahara}, \citenamefont {Glin\v{s}ek}, \citenamefont {Lee}, \citenamefont {Kim}, \citenamefont {Jeong}, \citenamefont {Kingon},\ and\ \citenamefont {Kim}}]{Won2019}%
  \BibitemOpen
  \bibfield  {author} {\bibinfo {author} {\bibfnamefont {S.~S.}\ \bibnamefont {Won}}, \bibinfo {author} {\bibfnamefont {H.}~\bibnamefont {Seo}}, \bibinfo {author} {\bibfnamefont {M.}~\bibnamefont {Kawahara}}, \bibinfo {author} {\bibfnamefont {S.}~\bibnamefont {Glin\v{s}ek}}, \bibinfo {author} {\bibfnamefont {J.}~\bibnamefont {Lee}}, \bibinfo {author} {\bibfnamefont {Y.}~\bibnamefont {Kim}}, \bibinfo {author} {\bibfnamefont {C.~K.}\ \bibnamefont {Jeong}}, \bibinfo {author} {\bibfnamefont {A.~I.}\ \bibnamefont {Kingon}}, \ and\ \bibinfo {author} {\bibfnamefont {S.-H.}\ \bibnamefont {Kim}},\ }\bibfield  {title} {\enquote {\bibinfo {title} {Flexible vibrational energy harvesting devices using strain-engineered perovskite piezoelectric thin films},}\ }\href {\doibase https://doi.org/10.1016/j.nanoen.2018.10.068} {\bibfield  {journal} {\bibinfo  {journal} {Nano Energy}\ }\textbf {\bibinfo {volume} {55}},\ \bibinfo {pages} {182--192} (\bibinfo {year} {2019})}\BibitemShut {NoStop}%
\bibitem [{\citenamefont {Yang}\ \emph {et~al.}(2019)\citenamefont {Yang}, \citenamefont {Han}, \citenamefont {Qian}, \citenamefont {Lv}, \citenamefont {Lin}, \citenamefont {Huang},\ and\ \citenamefont {Cheng}}]{YangBFMTO2019}%
  \BibitemOpen
  \bibfield  {author} {\bibinfo {author} {\bibfnamefont {C.}~\bibnamefont {Yang}}, \bibinfo {author} {\bibfnamefont {Y.}~\bibnamefont {Han}}, \bibinfo {author} {\bibfnamefont {J.}~\bibnamefont {Qian}}, \bibinfo {author} {\bibfnamefont {P.}~\bibnamefont {Lv}}, \bibinfo {author} {\bibfnamefont {X.}~\bibnamefont {Lin}}, \bibinfo {author} {\bibfnamefont {S.}~\bibnamefont {Huang}}, \ and\ \bibinfo {author} {\bibfnamefont {Z.}~\bibnamefont {Cheng}},\ }\bibfield  {title} {\enquote {\bibinfo {title} {Flexible, temperature-resistant, and fatigue-free ferroelectric memory based on {Bi(Fe$_{0.93}$Mn$_{0.05}$Ti$_{0.02}$)O$_3$} thin film},}\ }\href {\doibase 10.1021/acsami.9b01464} {\bibfield  {journal} {\bibinfo  {journal} {ACS Applied Materials \& Interfaces}\ }\textbf {\bibinfo {volume} {11}},\ \bibinfo {pages} {12647--12655} (\bibinfo {year} {2019})},\ \Eprint {http://arxiv.org/abs/https://doi.org/10.1021/acsami.9b01464} {https://doi.org/10.1021/acsami.9b01464} \BibitemShut {NoStop}%
\bibitem [{\citenamefont {Brennecka}\ \emph {et~al.}(2004)\citenamefont {Brennecka}, \citenamefont {Huebner}, \citenamefont {Tuttle},\ and\ \citenamefont {Clem}}]{Brennecka2004}%
  \BibitemOpen
  \bibfield  {author} {\bibinfo {author} {\bibfnamefont {G.~L.}\ \bibnamefont {Brennecka}}, \bibinfo {author} {\bibfnamefont {W.}~\bibnamefont {Huebner}}, \bibinfo {author} {\bibfnamefont {B.~A.}\ \bibnamefont {Tuttle}}, \ and\ \bibinfo {author} {\bibfnamefont {P.~G.}\ \bibnamefont {Clem}},\ }\bibfield  {title} {\enquote {\bibinfo {title} {Use of stress to produce highly oriented tetragonal lead zirconate titanate (pzt 40/60) thin films and resulting electrical properties},}\ }\href {\doibase https://doi.org/10.1111/j.1551-2916.2004.01459.x} {\bibfield  {journal} {\bibinfo  {journal} {Journal of the American Ceramic Society}\ }\textbf {\bibinfo {volume} {87}},\ \bibinfo {pages} {1459--1465} (\bibinfo {year} {2004})},\ \Eprint {http://arxiv.org/abs/https://ceramics.onlinelibrary.wiley.com/doi/pdf/10.1111/j.1551-2916.2004.01459.x} {https://ceramics.onlinelibrary.wiley.com/doi/pdf/10.1111/j.1551-2916.2004.01459.x} \BibitemShut {NoStop}%
\bibitem [{\citenamefont {Moreira}\ \emph {et~al.}(2024)\citenamefont {Moreira}, \citenamefont {Crêpellière}, \citenamefont {Polesel-Maris}, \citenamefont {Leturcq}, \citenamefont {Guillot}, \citenamefont {Fleming},\ and\ \citenamefont {Lunca-Popa}}]{Moreira2024}%
  \BibitemOpen
  \bibfield  {author} {\bibinfo {author} {\bibfnamefont {M.}~\bibnamefont {Moreira}}, \bibinfo {author} {\bibfnamefont {J.}~\bibnamefont {Crêpellière}}, \bibinfo {author} {\bibfnamefont {J.}~\bibnamefont {Polesel-Maris}}, \bibinfo {author} {\bibfnamefont {R.}~\bibnamefont {Leturcq}}, \bibinfo {author} {\bibfnamefont {J.}~\bibnamefont {Guillot}}, \bibinfo {author} {\bibfnamefont {Y.}~\bibnamefont {Fleming}}, \ and\ \bibinfo {author} {\bibfnamefont {P.}~\bibnamefont {Lunca-Popa}},\ }\bibfield  {title} {\enquote {\bibinfo {title} {Electrical properties of strained off-stoichiometric {Cu}–{Cr}–{O} delafossite thin films},}\ }\href {\doibase 10.1088/1361-648X/ad2a07} {\bibfield  {journal} {\bibinfo  {journal} {Journal of Physics Condensed Matter}\ }\textbf {\bibinfo {volume} {36}} (\bibinfo {year} {2024}),\ 10.1088/1361-648X/ad2a07}\BibitemShut {NoStop}%
\bibitem [{\citenamefont {Schenk}\ \emph {et~al.}(2021)\citenamefont {Schenk}, \citenamefont {Bencan}, \citenamefont {Drazic}, \citenamefont {Condurache}, \citenamefont {Valle}, \citenamefont {Adib}, \citenamefont {Aruchamy}, \citenamefont {Granzow}, \citenamefont {Defay},\ and\ \citenamefont {Glinsek}}]{Schenk2021APL}%
  \BibitemOpen
  \bibfield  {author} {\bibinfo {author} {\bibfnamefont {T.}~\bibnamefont {Schenk}}, \bibinfo {author} {\bibfnamefont {A.}~\bibnamefont {Bencan}}, \bibinfo {author} {\bibfnamefont {G.}~\bibnamefont {Drazic}}, \bibinfo {author} {\bibfnamefont {O.}~\bibnamefont {Condurache}}, \bibinfo {author} {\bibfnamefont {N.}~\bibnamefont {Valle}}, \bibinfo {author} {\bibfnamefont {B.~E.}\ \bibnamefont {Adib}}, \bibinfo {author} {\bibfnamefont {N.}~\bibnamefont {Aruchamy}}, \bibinfo {author} {\bibfnamefont {T.}~\bibnamefont {Granzow}}, \bibinfo {author} {\bibfnamefont {E.}~\bibnamefont {Defay}}, \ and\ \bibinfo {author} {\bibfnamefont {S.}~\bibnamefont {Glinsek}},\ }\bibfield  {title} {\enquote {\bibinfo {title} {Enhancement of ferroelectricity and orientation in solution-derived hafnia thin films through heterogeneous grain nucleation},}\ }\href {\doibase 10.1063/5.0045966} {\bibfield  {journal} {\bibinfo  {journal} {Applied Physics Letters}\ ,\ \bibinfo {pages} {162902}} (\bibinfo {year} {2021})},\ \bibinfo {note} {aIP
  Publishing}\BibitemShut {NoStop}%
\bibitem [{\citenamefont {Coleman}\ \emph {et~al.}(2019)\citenamefont {Coleman}, \citenamefont {Walker}, \citenamefont {Beechem},\ and\ \citenamefont {Trolier-Mckinstry}}]{Coleman2019}%
  \BibitemOpen
  \bibfield  {author} {\bibinfo {author} {\bibfnamefont {K.}~\bibnamefont {Coleman}}, \bibinfo {author} {\bibfnamefont {J.}~\bibnamefont {Walker}}, \bibinfo {author} {\bibfnamefont {T.}~\bibnamefont {Beechem}}, \ and\ \bibinfo {author} {\bibfnamefont {S.}~\bibnamefont {Trolier-Mckinstry}},\ }\bibfield  {title} {\enquote {\bibinfo {title} {Effect of stresses on the dielectric and piezoelectric properties of {Pb(Zr$_{0.52}$Ti$_{0.48}$)O$_3$} thin films},}\ }\href {\doibase 10.1063/1.5095765} {\bibfield  {journal} {\bibinfo  {journal} {Journal of Applied Physics}\ }\textbf {\bibinfo {volume} {126}} (\bibinfo {year} {2019}),\ 10.1063/1.5095765}\BibitemShut {NoStop}%
\bibitem [{\citenamefont {Imre}(2007)}]{Imre2007-NegativeP}%
  \BibitemOpen
  \bibfield  {author} {\bibinfo {author} {\bibfnamefont {A.~R.}\ \bibnamefont {Imre}},\ }\bibfield  {title} {\enquote {\bibinfo {title} {On the existence of negative pressure states},}\ }\href {\doibase 10.1002/pssb.200572708} {\bibfield  {journal} {\bibinfo  {journal} {physica status solidi (b)}\ }\textbf {\bibinfo {volume} {244}},\ \bibinfo {pages} {893--899} (\bibinfo {year} {2007})}\BibitemShut {NoStop}%
\bibitem [{\citenamefont {Liu}\ \emph {et~al.}(2009)\citenamefont {Liu}, \citenamefont {Ni}, \citenamefont {Ren}, \citenamefont {Xu}, \citenamefont {Song},\ and\ \citenamefont {Han}}]{liu_negative_2009}%
  \BibitemOpen
  \bibfield  {author} {\bibinfo {author} {\bibfnamefont {Y.}~\bibnamefont {Liu}}, \bibinfo {author} {\bibfnamefont {L.}~\bibnamefont {Ni}}, \bibinfo {author} {\bibfnamefont {Z.}~\bibnamefont {Ren}}, \bibinfo {author} {\bibfnamefont {G.}~\bibnamefont {Xu}}, \bibinfo {author} {\bibfnamefont {C.}~\bibnamefont {Song}}, \ and\ \bibinfo {author} {\bibfnamefont {G.}~\bibnamefont {Han}},\ }\bibfield  {title} {{\selectlanguage {en}\enquote {\bibinfo {title} {Negative pressure induced ferroelectric phase transition in rutile {TiO$_2$}},}\ }}\href {\doibase 10.1088/0953-8984/21/27/275901} {\bibfield  {journal} {\bibinfo  {journal} {Journal of Physics: Condensed Matter}\ }\textbf {\bibinfo {volume} {21}},\ \bibinfo {pages} {275901} (\bibinfo {year} {2009})}\BibitemShut {NoStop}%
\bibitem [{\citenamefont {Tinte}, \citenamefont {Rabe},\ and\ \citenamefont {Vanderbilt}(2003)}]{tinte_anomalous_2003}%
  \BibitemOpen
  \bibfield  {author} {\bibinfo {author} {\bibfnamefont {S.}~\bibnamefont {Tinte}}, \bibinfo {author} {\bibfnamefont {K.~M.}\ \bibnamefont {Rabe}}, \ and\ \bibinfo {author} {\bibfnamefont {D.}~\bibnamefont {Vanderbilt}},\ }\bibfield  {title} {\enquote {\bibinfo {title} {Anomalous enhancement of tetragonality in {PbTiO$_3$} induced by negative pressure},}\ }\href {\doibase 10.1103/PhysRevB.68.144105} {\bibfield  {journal} {\bibinfo  {journal} {Physical Review B}\ }\textbf {\bibinfo {volume} {68}},\ \bibinfo {pages} {144105} (\bibinfo {year} {2003})}\BibitemShut {NoStop}%
\bibitem [{\citenamefont {Aligia}\ \emph {et~al.}(2001)\citenamefont {Aligia}, \citenamefont {Petrone}, \citenamefont {Sofo},\ and\ \citenamefont {Alascio}}]{Aligia2001}%
  \BibitemOpen
  \bibfield  {author} {\bibinfo {author} {\bibfnamefont {A.~A.}\ \bibnamefont {Aligia}}, \bibinfo {author} {\bibfnamefont {P.}~\bibnamefont {Petrone}}, \bibinfo {author} {\bibfnamefont {J.~O.}\ \bibnamefont {Sofo}}, \ and\ \bibinfo {author} {\bibfnamefont {B.}~\bibnamefont {Alascio}},\ }\bibfield  {title} {\enquote {\bibinfo {title} {Metal-insulator transition in the double perovskites},}\ }\href {\doibase 10.1103/PhysRevB.64.092414} {\bibfield  {journal} {\bibinfo  {journal} {Physical Review B}\ }\textbf {\bibinfo {volume} {64}},\ \bibinfo {pages} {092414} (\bibinfo {year} {2001})}\BibitemShut {NoStop}%
\bibitem [{\citenamefont {Sharma}\ \emph {et~al.}(2017)\citenamefont {Sharma}, \citenamefont {Herklotz}, \citenamefont {Ward},\ and\ \citenamefont {Reboredo}}]{sharma_designing_2017}%
  \BibitemOpen
  \bibfield  {author} {\bibinfo {author} {\bibfnamefont {V.}~\bibnamefont {Sharma}}, \bibinfo {author} {\bibfnamefont {A.}~\bibnamefont {Herklotz}}, \bibinfo {author} {\bibfnamefont {T.~Z.}\ \bibnamefont {Ward}}, \ and\ \bibinfo {author} {\bibfnamefont {F.~A.}\ \bibnamefont {Reboredo}},\ }\bibfield  {title} {{\selectlanguage {en}\enquote {\bibinfo {title} {Designing functionality in perovskite thin films using ion implantation techniques: {Assessment} and insights from first-principles calculations},}\ }}\href {\doibase 10.1038/s41598-017-11158-4} {\bibfield  {journal} {\bibinfo  {journal} {Scientific Reports}\ }\textbf {\bibinfo {volume} {7}},\ \bibinfo {pages} {1--10} (\bibinfo {year} {2017})},\ \bibinfo {note} {number: 1 Publisher: Nature Publishing Group}\BibitemShut {NoStop}%
\bibitem [{\citenamefont {Barazani}\ \emph {et~al.}(2023)\citenamefont {Barazani}, \citenamefont {Das}, \citenamefont {Huang}, \citenamefont {Rakshit}, \citenamefont {Saguy}, \citenamefont {Salev}, \citenamefont {del Valle}, \citenamefont {Toroker}, \citenamefont {Schuller},\ and\ \citenamefont {Kalcheim}}]{Barazani2023}%
  \BibitemOpen
  \bibfield  {author} {\bibinfo {author} {\bibfnamefont {E.}~\bibnamefont {Barazani}}, \bibinfo {author} {\bibfnamefont {D.}~\bibnamefont {Das}}, \bibinfo {author} {\bibfnamefont {C.}~\bibnamefont {Huang}}, \bibinfo {author} {\bibfnamefont {A.}~\bibnamefont {Rakshit}}, \bibinfo {author} {\bibfnamefont {C.}~\bibnamefont {Saguy}}, \bibinfo {author} {\bibfnamefont {P.}~\bibnamefont {Salev}}, \bibinfo {author} {\bibfnamefont {J.}~\bibnamefont {del Valle}}, \bibinfo {author} {\bibfnamefont {M.~C.}\ \bibnamefont {Toroker}}, \bibinfo {author} {\bibfnamefont {I.~K.}\ \bibnamefont {Schuller}}, \ and\ \bibinfo {author} {\bibfnamefont {Y.}~\bibnamefont {Kalcheim}},\ }\bibfield  {title} {\enquote {\bibinfo {title} {Positive and negative pressure regimes in anisotropically strained {V$_2$O$_3$} films},}\ }\href {\doibase 10.1002/adfm.202211801} {\bibfield  {journal} {\bibinfo  {journal} {Advanced Functional Materials}\ }\textbf {\bibinfo {volume} {33}} (\bibinfo {year} {2023}),\ 10.1002/adfm.202211801}\BibitemShut
  {NoStop}%
\bibitem [{\citenamefont {Aguirre-Tostado}\ \emph {et~al.}(2004)\citenamefont {Aguirre-Tostado}, \citenamefont {Herrera-Gómez}, \citenamefont {Woicik}, \citenamefont {Droopad}, \citenamefont {Yu}, \citenamefont {Schlom}, \citenamefont {Karapetrova}, \citenamefont {Zschack},\ and\ \citenamefont {Pianetta}}]{Aguirre2004}%
  \BibitemOpen
  \bibfield  {author} {\bibinfo {author} {\bibfnamefont {F.~S.}\ \bibnamefont {Aguirre-Tostado}}, \bibinfo {author} {\bibfnamefont {A.}~\bibnamefont {Herrera-Gómez}}, \bibinfo {author} {\bibfnamefont {J.~C.}\ \bibnamefont {Woicik}}, \bibinfo {author} {\bibfnamefont {R.}~\bibnamefont {Droopad}}, \bibinfo {author} {\bibfnamefont {Z.}~\bibnamefont {Yu}}, \bibinfo {author} {\bibfnamefont {D.~G.}\ \bibnamefont {Schlom}}, \bibinfo {author} {\bibfnamefont {J.}~\bibnamefont {Karapetrova}}, \bibinfo {author} {\bibfnamefont {P.}~\bibnamefont {Zschack}}, \ and\ \bibinfo {author} {\bibfnamefont {P.}~\bibnamefont {Pianetta}},\ }\bibfield  {title} {\enquote {\bibinfo {title} {Displacive phase transition in {SrTiO$_3$} thin films grown on {S}i(001)},}\ }\href {\doibase 10.1116/1.1765657} {\bibfield  {journal} {\bibinfo  {journal} {Journal of Vacuum Science \& Technology A: Vacuum, Surfaces, and Films}\ }\textbf {\bibinfo {volume} {22}},\ \bibinfo {pages} {1356--1360} (\bibinfo {year} {2004})}\BibitemShut {NoStop}%
\bibitem [{\citenamefont {Guo}\ \emph {et~al.}(2015)\citenamefont {Guo}, \citenamefont {Dong}, \citenamefont {Rack}, \citenamefont {Budai}, \citenamefont {Beekman}, \citenamefont {Gai}, \citenamefont {Siemons}, \citenamefont {Gonzalez}, \citenamefont {Timilsina}, \citenamefont {Wong}, \citenamefont {Herklotz}, \citenamefont {Snijders}, \citenamefont {Dagotto},\ and\ \citenamefont {Ward}}]{guo_strain_2015}%
  \BibitemOpen
  \bibfield  {author} {\bibinfo {author} {\bibfnamefont {H.}~\bibnamefont {Guo}}, \bibinfo {author} {\bibfnamefont {S.}~\bibnamefont {Dong}}, \bibinfo {author} {\bibfnamefont {P.~D.}\ \bibnamefont {Rack}}, \bibinfo {author} {\bibfnamefont {J.~D.}\ \bibnamefont {Budai}}, \bibinfo {author} {\bibfnamefont {C.}~\bibnamefont {Beekman}}, \bibinfo {author} {\bibfnamefont {Z.}~\bibnamefont {Gai}}, \bibinfo {author} {\bibfnamefont {W.}~\bibnamefont {Siemons}}, \bibinfo {author} {\bibfnamefont {C.~M.}\ \bibnamefont {Gonzalez}}, \bibinfo {author} {\bibfnamefont {R.}~\bibnamefont {Timilsina}}, \bibinfo {author} {\bibfnamefont {A.~T.}\ \bibnamefont {Wong}}, \bibinfo {author} {\bibfnamefont {A.}~\bibnamefont {Herklotz}}, \bibinfo {author} {\bibfnamefont {P.~C.}\ \bibnamefont {Snijders}}, \bibinfo {author} {\bibfnamefont {E.}~\bibnamefont {Dagotto}}, \ and\ \bibinfo {author} {\bibfnamefont {T.~Z.}\ \bibnamefont {Ward}},\ }\bibfield  {title} {\enquote {\bibinfo {title} {Strain {Doping}: {Reversible} {Single}-{Axis} {Control}
  of a {Complex} {Oxide} {Lattice} via {Helium} {Implantation}},}\ }\href {\doibase 10.1103/PhysRevLett.114.256801} {\bibfield  {journal} {\bibinfo  {journal} {Physical Review Letters}\ }\textbf {\bibinfo {volume} {114}},\ \bibinfo {pages} {256801} (\bibinfo {year} {2015})}\BibitemShut {NoStop}%
\bibitem [{\citenamefont {Toulouse}\ \emph {et~al.}(2021)\citenamefont {Toulouse}, \citenamefont {Fischer}, \citenamefont {Farokhipoor}, \citenamefont {Yedra}, \citenamefont {Carlà}, \citenamefont {Jarnac}, \citenamefont {Elkaim}, \citenamefont {Fertey}, \citenamefont {Audinot}, \citenamefont {Wirtz}, \citenamefont {Noheda}, \citenamefont {Garcia}, \citenamefont {Fusil}, \citenamefont {Alonso}, \citenamefont {Guennou},\ and\ \citenamefont {Kreisel}}]{toulouse_patterning_2021}%
  \BibitemOpen
  \bibfield  {author} {\bibinfo {author} {\bibfnamefont {C.}~\bibnamefont {Toulouse}}, \bibinfo {author} {\bibfnamefont {J.}~\bibnamefont {Fischer}}, \bibinfo {author} {\bibfnamefont {S.}~\bibnamefont {Farokhipoor}}, \bibinfo {author} {\bibfnamefont {L.}~\bibnamefont {Yedra}}, \bibinfo {author} {\bibfnamefont {F.}~\bibnamefont {Carlà}}, \bibinfo {author} {\bibfnamefont {A.}~\bibnamefont {Jarnac}}, \bibinfo {author} {\bibfnamefont {E.}~\bibnamefont {Elkaim}}, \bibinfo {author} {\bibfnamefont {P.}~\bibnamefont {Fertey}}, \bibinfo {author} {\bibfnamefont {J.-N.}\ \bibnamefont {Audinot}}, \bibinfo {author} {\bibfnamefont {T.}~\bibnamefont {Wirtz}}, \bibinfo {author} {\bibfnamefont {B.}~\bibnamefont {Noheda}}, \bibinfo {author} {\bibfnamefont {V.}~\bibnamefont {Garcia}}, \bibinfo {author} {\bibfnamefont {S.}~\bibnamefont {Fusil}}, \bibinfo {author} {\bibfnamefont {I.~P.}\ \bibnamefont {Alonso}}, \bibinfo {author} {\bibfnamefont {M.}~\bibnamefont {Guennou}}, \ and\ \bibinfo {author} {\bibfnamefont
  {J.}~\bibnamefont {Kreisel}},\ }\bibfield  {title} {\enquote {\bibinfo {title} {Patterning enhanced tetragonality in {BiFeO$_3$} thin films with effective negative pressure by helium implantation},}\ }\href {\doibase 10.1103/PhysRevMaterials.5.024404} {\bibfield  {journal} {\bibinfo  {journal} {Physical Review Materials}\ }\textbf {\bibinfo {volume} {5}},\ \bibinfo {pages} {024404} (\bibinfo {year} {2021})},\ \bibinfo {note} {publisher: American Physical Society}\BibitemShut {NoStop}%
\bibitem [{\citenamefont {Livengood}\ \emph {et~al.}(2009)\citenamefont {Livengood}, \citenamefont {Tan}, \citenamefont {Greenzweig}, \citenamefont {Notte},\ and\ \citenamefont {McVey}}]{livengood_subsurface_2009}%
  \BibitemOpen
  \bibfield  {author} {\bibinfo {author} {\bibfnamefont {R.}~\bibnamefont {Livengood}}, \bibinfo {author} {\bibfnamefont {S.}~\bibnamefont {Tan}}, \bibinfo {author} {\bibfnamefont {Y.}~\bibnamefont {Greenzweig}}, \bibinfo {author} {\bibfnamefont {J.}~\bibnamefont {Notte}}, \ and\ \bibinfo {author} {\bibfnamefont {S.}~\bibnamefont {McVey}},\ }\bibfield  {title} {\enquote {\bibinfo {title} {Subsurface damage from helium ions as a function of dose, beam energy, and dose rate},}\ }\href {\doibase 10.1116/1.3237101} {\bibfield  {journal} {\bibinfo  {journal} {Journal of Vacuum Science \& Technology: Part B-Microelectronics \& Nanometer Structures}\ }\textbf {\bibinfo {volume} {27}},\ \bibinfo {pages} {3244--3249} (\bibinfo {year} {2009})}\BibitemShut {NoStop}%
\bibitem [{\citenamefont {Heron}\ \emph {et~al.}(2011)\citenamefont {Heron}, \citenamefont {Trassin}, \citenamefont {Ashraf}, \citenamefont {Gajek}, \citenamefont {He}, \citenamefont {Yang}, \citenamefont {Nikonov}, \citenamefont {Chu}, \citenamefont {Salahuddin},\ and\ \citenamefont {Ramesh}}]{Heron2011}%
  \BibitemOpen
  \bibfield  {author} {\bibinfo {author} {\bibfnamefont {J.~T.}\ \bibnamefont {Heron}}, \bibinfo {author} {\bibfnamefont {M.}~\bibnamefont {Trassin}}, \bibinfo {author} {\bibfnamefont {K.}~\bibnamefont {Ashraf}}, \bibinfo {author} {\bibfnamefont {M.}~\bibnamefont {Gajek}}, \bibinfo {author} {\bibfnamefont {Q.}~\bibnamefont {He}}, \bibinfo {author} {\bibfnamefont {S.~Y.}\ \bibnamefont {Yang}}, \bibinfo {author} {\bibfnamefont {D.~E.}\ \bibnamefont {Nikonov}}, \bibinfo {author} {\bibfnamefont {Y.-H.}\ \bibnamefont {Chu}}, \bibinfo {author} {\bibfnamefont {S.}~\bibnamefont {Salahuddin}}, \ and\ \bibinfo {author} {\bibfnamefont {R.}~\bibnamefont {Ramesh}},\ }\bibfield  {title} {\enquote {\bibinfo {title} {Electric-field-induced magnetization reversal in a ferromagnet-multiferroic heterostructure},}\ }\href {\doibase 10.1103/PhysRevLett.107.217202} {\bibfield  {journal} {\bibinfo  {journal} {Physical Review Letters}\ }\textbf {\bibinfo {volume} {107}},\ \bibinfo {pages} {217202} (\bibinfo {year}
  {2011})}\BibitemShut {NoStop}%
\bibitem [{\citenamefont {Lebeugle}\ \emph {et~al.}(2007{\natexlab{a}})\citenamefont {Lebeugle}, \citenamefont {Colson}, \citenamefont {Forget},\ and\ \citenamefont {Viret}}]{lebeugle_very_2007}%
  \BibitemOpen
  \bibfield  {author} {\bibinfo {author} {\bibfnamefont {D.}~\bibnamefont {Lebeugle}}, \bibinfo {author} {\bibfnamefont {D.}~\bibnamefont {Colson}}, \bibinfo {author} {\bibfnamefont {A.}~\bibnamefont {Forget}}, \ and\ \bibinfo {author} {\bibfnamefont {M.}~\bibnamefont {Viret}},\ }\bibfield  {title} {\enquote {\bibinfo {title} {Very large spontaneous electric polarization in {BiFeO$_3$} single crystals at room temperature and its evolution under cycling fields},}\ }\href {\doibase 10.1063/1.2753390} {\bibfield  {journal} {\bibinfo  {journal} {Applied Physics Letters}\ }\textbf {\bibinfo {volume} {91}},\ \bibinfo {pages} {022907} (\bibinfo {year} {2007}{\natexlab{a}})},\ \bibinfo {note} {publisher: American Institute of Physics}\BibitemShut {NoStop}%
\bibitem [{\citenamefont {Lebeugle}\ \emph {et~al.}(2007{\natexlab{b}})\citenamefont {Lebeugle}, \citenamefont {Colson}, \citenamefont {Forget}, \citenamefont {Viret}, \citenamefont {Bonville}, \citenamefont {Marucco},\ and\ \citenamefont {Fusil}}]{lebeugle_room-temperature_2007}%
  \BibitemOpen
  \bibfield  {author} {\bibinfo {author} {\bibfnamefont {D.}~\bibnamefont {Lebeugle}}, \bibinfo {author} {\bibfnamefont {D.}~\bibnamefont {Colson}}, \bibinfo {author} {\bibfnamefont {A.}~\bibnamefont {Forget}}, \bibinfo {author} {\bibfnamefont {M.}~\bibnamefont {Viret}}, \bibinfo {author} {\bibfnamefont {P.}~\bibnamefont {Bonville}}, \bibinfo {author} {\bibfnamefont {J.~F.}\ \bibnamefont {Marucco}}, \ and\ \bibinfo {author} {\bibfnamefont {S.}~\bibnamefont {Fusil}},\ }\bibfield  {title} {\enquote {\bibinfo {title} {Room-temperature coexistence of large electric polarization and magnetic order in {BiFeO$_3$} single crystals},}\ }\href {\doibase 10.1103/PhysRevB.76.024116} {\bibfield  {journal} {\bibinfo  {journal} {Physical Review B}\ }\textbf {\bibinfo {volume} {76}},\ \bibinfo {pages} {024116} (\bibinfo {year} {2007}{\natexlab{b}})}\BibitemShut {NoStop}%
\bibitem [{\citenamefont {Sando}\ \emph {et~al.}(2016)\citenamefont {Sando}, \citenamefont {Xu}, \citenamefont {Bellaiche},\ and\ \citenamefont {Nagarajan}}]{sando_multiferroic_2016}%
  \BibitemOpen
  \bibfield  {author} {\bibinfo {author} {\bibfnamefont {D.}~\bibnamefont {Sando}}, \bibinfo {author} {\bibfnamefont {B.}~\bibnamefont {Xu}}, \bibinfo {author} {\bibfnamefont {L.}~\bibnamefont {Bellaiche}}, \ and\ \bibinfo {author} {\bibfnamefont {V.}~\bibnamefont {Nagarajan}},\ }\bibfield  {title} {\enquote {\bibinfo {title} {A multiferroic on the brink: {Uncovering} the nuances of strain-induced transitions in {BiFeO$_3$}},}\ }\href {\doibase 10.1063/1.4944558} {\bibfield  {journal} {\bibinfo  {journal} {Applied Physics Reviews}\ }\textbf {\bibinfo {volume} {3}},\ \bibinfo {pages} {011106} (\bibinfo {year} {2016})}\BibitemShut {NoStop}%
\bibitem [{\citenamefont {Herklotz}\ \emph {et~al.}(2019{\natexlab{a}})\citenamefont {Herklotz}, \citenamefont {Rus}, \citenamefont {Balke}, \citenamefont {Rouleau}, \citenamefont {Guo}, \citenamefont {Huon}, \citenamefont {KC}, \citenamefont {Roth}, \citenamefont {Yang}, \citenamefont {Vaswani}, \citenamefont {Wang}, \citenamefont {Orth}, \citenamefont {Scheurer},\ and\ \citenamefont {Ward}}]{herklotz_designing_2019}%
  \BibitemOpen
  \bibfield  {author} {\bibinfo {author} {\bibfnamefont {A.}~\bibnamefont {Herklotz}}, \bibinfo {author} {\bibfnamefont {S.~F.}\ \bibnamefont {Rus}}, \bibinfo {author} {\bibfnamefont {N.}~\bibnamefont {Balke}}, \bibinfo {author} {\bibfnamefont {C.}~\bibnamefont {Rouleau}}, \bibinfo {author} {\bibfnamefont {E.-J.}\ \bibnamefont {Guo}}, \bibinfo {author} {\bibfnamefont {A.}~\bibnamefont {Huon}}, \bibinfo {author} {\bibfnamefont {S.}~\bibnamefont {KC}}, \bibinfo {author} {\bibfnamefont {R.}~\bibnamefont {Roth}}, \bibinfo {author} {\bibfnamefont {X.}~\bibnamefont {Yang}}, \bibinfo {author} {\bibfnamefont {C.}~\bibnamefont {Vaswani}}, \bibinfo {author} {\bibfnamefont {J.}~\bibnamefont {Wang}}, \bibinfo {author} {\bibfnamefont {P.~P.}\ \bibnamefont {Orth}}, \bibinfo {author} {\bibfnamefont {M.~S.}\ \bibnamefont {Scheurer}}, \ and\ \bibinfo {author} {\bibfnamefont {T.~Z.}\ \bibnamefont {Ward}},\ }\bibfield  {title} {\enquote {\bibinfo {title} {Designing {Morphotropic} {Phase} {Composition} in {BiFeO$_3$}},}\ }\href
  {\doibase 10.1021/acs.nanolett.8b04322} {\bibfield  {journal} {\bibinfo  {journal} {Nano Letters}\ } (\bibinfo {year} {2019}{\natexlab{a}}),\ 10.1021/acs.nanolett.8b04322}\BibitemShut {NoStop}%
\bibitem [{\citenamefont {Chen}\ \emph {et~al.}(2019)\citenamefont {Chen}, \citenamefont {Wang}, \citenamefont {Cai}, \citenamefont {Xu}, \citenamefont {Li}, \citenamefont {Zhou}, \citenamefont {Luo}, \citenamefont {Fan}, \citenamefont {Qin}, \citenamefont {Zeng}, \citenamefont {Lu}, \citenamefont {Gao}, \citenamefont {Kentsch}, \citenamefont {Yang}, \citenamefont {Zhou}, \citenamefont {Wang}, \citenamefont {Zhu}, \citenamefont {Zhou}, \citenamefont {Chen},\ and\ \citenamefont {Liu}}]{Chen2019}%
  \BibitemOpen
  \bibfield  {author} {\bibinfo {author} {\bibfnamefont {C.}~\bibnamefont {Chen}}, \bibinfo {author} {\bibfnamefont {C.}~\bibnamefont {Wang}}, \bibinfo {author} {\bibfnamefont {X.}~\bibnamefont {Cai}}, \bibinfo {author} {\bibfnamefont {C.}~\bibnamefont {Xu}}, \bibinfo {author} {\bibfnamefont {C.}~\bibnamefont {Li}}, \bibinfo {author} {\bibfnamefont {J.}~\bibnamefont {Zhou}}, \bibinfo {author} {\bibfnamefont {Z.}~\bibnamefont {Luo}}, \bibinfo {author} {\bibfnamefont {Z.}~\bibnamefont {Fan}}, \bibinfo {author} {\bibfnamefont {M.}~\bibnamefont {Qin}}, \bibinfo {author} {\bibfnamefont {M.}~\bibnamefont {Zeng}}, \bibinfo {author} {\bibfnamefont {X.}~\bibnamefont {Lu}}, \bibinfo {author} {\bibfnamefont {X.}~\bibnamefont {Gao}}, \bibinfo {author} {\bibfnamefont {U.}~\bibnamefont {Kentsch}}, \bibinfo {author} {\bibfnamefont {P.}~\bibnamefont {Yang}}, \bibinfo {author} {\bibfnamefont {G.}~\bibnamefont {Zhou}}, \bibinfo {author} {\bibfnamefont {N.}~\bibnamefont {Wang}}, \bibinfo {author} {\bibfnamefont
  {Y.}~\bibnamefont {Zhu}}, \bibinfo {author} {\bibfnamefont {S.}~\bibnamefont {Zhou}}, \bibinfo {author} {\bibfnamefont {D.}~\bibnamefont {Chen}}, \ and\ \bibinfo {author} {\bibfnamefont {J.-M.}\ \bibnamefont {Liu}},\ }\bibfield  {title} {\enquote {\bibinfo {title} {Controllable defect driven symmetry change and domain structure evolution in {BiFeO$_3$} with enhanced tetragonality},}\ }\href {\doibase 10.1039/C9NR00932A} {\bibfield  {journal} {\bibinfo  {journal} {Nanoscale}\ }\textbf {\bibinfo {volume} {11}},\ \bibinfo {pages} {8110--8118} (\bibinfo {year} {2019})}\BibitemShut {NoStop}%
\bibitem [{\citenamefont {Cai}\ \emph {et~al.}(2023)\citenamefont {Cai}, \citenamefont {Chen}, \citenamefont {Xie}, \citenamefont {Wang}, \citenamefont {Gui}, \citenamefont {Gao}, \citenamefont {Kentsch}, \citenamefont {Zhou}, \citenamefont {Gao}, \citenamefont {Chen}, \citenamefont {Zhou}, \citenamefont {Gao}, \citenamefont {Liu}, \citenamefont {Zhu},\ and\ \citenamefont {Chen}}]{Cai2023}%
  \BibitemOpen
  \bibfield  {author} {\bibinfo {author} {\bibfnamefont {X.}~\bibnamefont {Cai}}, \bibinfo {author} {\bibfnamefont {C.}~\bibnamefont {Chen}}, \bibinfo {author} {\bibfnamefont {L.}~\bibnamefont {Xie}}, \bibinfo {author} {\bibfnamefont {C.}~\bibnamefont {Wang}}, \bibinfo {author} {\bibfnamefont {Z.}~\bibnamefont {Gui}}, \bibinfo {author} {\bibfnamefont {Y.}~\bibnamefont {Gao}}, \bibinfo {author} {\bibfnamefont {U.}~\bibnamefont {Kentsch}}, \bibinfo {author} {\bibfnamefont {G.}~\bibnamefont {Zhou}}, \bibinfo {author} {\bibfnamefont {X.}~\bibnamefont {Gao}}, \bibinfo {author} {\bibfnamefont {Y.}~\bibnamefont {Chen}}, \bibinfo {author} {\bibfnamefont {S.}~\bibnamefont {Zhou}}, \bibinfo {author} {\bibfnamefont {W.}~\bibnamefont {Gao}}, \bibinfo {author} {\bibfnamefont {J.-M.}\ \bibnamefont {Liu}}, \bibinfo {author} {\bibfnamefont {Y.}~\bibnamefont {Zhu}}, \ and\ \bibinfo {author} {\bibfnamefont {D.}~\bibnamefont {Chen}},\ }\bibfield  {title} {\enquote {\bibinfo {title} {In-plane charged antiphase boundary and 180°
  domain wall in a ferroelectric film},}\ }\href {\doibase 10.1038/s41467-023-44091-4} {\bibfield  {journal} {\bibinfo  {journal} {Nature Communications}\ }\textbf {\bibinfo {volume} {14}},\ \bibinfo {pages} {8174} (\bibinfo {year} {2023})}\BibitemShut {NoStop}%
\bibitem [{\citenamefont {Mart\'inez}\ \emph {et~al.}(2023)\citenamefont {Mart\'inez}, \citenamefont {Grysan}, \citenamefont {Girod}, \citenamefont {Kovacova}, \citenamefont {Glin\v{s}ek},\ and\ \citenamefont {Granzow}}]{BlazquezMartinez2023}%
  \BibitemOpen
  \bibfield  {author} {\bibinfo {author} {\bibfnamefont {A.~B.}\ \bibnamefont {Mart\'inez}}, \bibinfo {author} {\bibfnamefont {P.}~\bibnamefont {Grysan}}, \bibinfo {author} {\bibfnamefont {S.}~\bibnamefont {Girod}}, \bibinfo {author} {\bibfnamefont {V.}~\bibnamefont {Kovacova}}, \bibinfo {author} {\bibfnamefont {S.}~\bibnamefont {Glin\v{s}ek}}, \ and\ \bibinfo {author} {\bibfnamefont {T.}~\bibnamefont {Granzow}},\ }\bibfield  {title} {\enquote {\bibinfo {title} {Stress-tuning the bulk photovoltaic response in polycrystalline bismuth ferrite films},}\ }\href {\doibase 10.1063/5.0136800} {\bibfield  {journal} {\bibinfo  {journal} {Applied Physics Letters}\ }\textbf {\bibinfo {volume} {122}},\ \bibinfo {pages} {152903} (\bibinfo {year} {2023})}\BibitemShut {NoStop}%
\bibitem [{\citenamefont {Bl\'azquez~Mart\'inez}\ \emph {et~al.}(2021)\citenamefont {Bl\'azquez~Mart\'inez}, \citenamefont {Godard}, \citenamefont {Aruchamy}, \citenamefont {Milesi-Brault}, \citenamefont {Condurache}, \citenamefont {Bencan}, \citenamefont {Glin\v{s}ek},\ and\ \citenamefont {Granzow}}]{BlazquezMartinez2021}%
  \BibitemOpen
  \bibfield  {author} {\bibinfo {author} {\bibfnamefont {A.}~\bibnamefont {Bl\'azquez~Mart\'inez}}, \bibinfo {author} {\bibfnamefont {N.}~\bibnamefont {Godard}}, \bibinfo {author} {\bibfnamefont {N.}~\bibnamefont {Aruchamy}}, \bibinfo {author} {\bibfnamefont {C.}~\bibnamefont {Milesi-Brault}}, \bibinfo {author} {\bibfnamefont {O.}~\bibnamefont {Condurache}}, \bibinfo {author} {\bibfnamefont {A.}~\bibnamefont {Bencan}}, \bibinfo {author} {\bibfnamefont {S.}~\bibnamefont {Glin\v{s}ek}}, \ and\ \bibinfo {author} {\bibfnamefont {T.}~\bibnamefont {Granzow}},\ }\bibfield  {title} {\enquote {\bibinfo {title} {{Solution-processed \ch{BiFeO3} thin films with low leakage current}},}\ }\href {\doibase 10.1016/j.jeurceramsoc.2021.05.051} {\bibfield  {journal} {\bibinfo  {journal} {Journal of the European Ceramic Society}\ }\textbf {\bibinfo {volume} {41}},\ \bibinfo {pages} {6449--6455} (\bibinfo {year} {2021})}\BibitemShut {NoStop}%
\bibitem [{\citenamefont {Qi}\ \emph {et~al.}(2005)\citenamefont {Qi}, \citenamefont {Dho}, \citenamefont {Tomov}, \citenamefont {Blamire},\ and\ \citenamefont {MacManus-Driscoll}}]{Qi2005GreatlyBiFeO3}%
  \BibitemOpen
  \bibfield  {author} {\bibinfo {author} {\bibfnamefont {X.}~\bibnamefont {Qi}}, \bibinfo {author} {\bibfnamefont {J.}~\bibnamefont {Dho}}, \bibinfo {author} {\bibfnamefont {R.}~\bibnamefont {Tomov}}, \bibinfo {author} {\bibfnamefont {M.~G.}\ \bibnamefont {Blamire}}, \ and\ \bibinfo {author} {\bibfnamefont {J.~L.}\ \bibnamefont {MacManus-Driscoll}},\ }\bibfield  {title} {\enquote {\bibinfo {title} {{Greatly reduced leakage current and conduction mechanism in aliovalent-ion-doped \ch{BiFeO3}}},}\ }\href {\doibase 10.1063/1.1862336} {\bibfield  {journal} {\bibinfo  {journal} {Applied Physics Letters}\ }\textbf {\bibinfo {volume} {86}},\ \bibinfo {pages} {062903} (\bibinfo {year} {2005})}\BibitemShut {NoStop}%
\bibitem [{\citenamefont {Singh}\ \emph {et~al.}(2007)\citenamefont {Singh}, \citenamefont {Ishiwara}, \citenamefont {Sato},\ and\ \citenamefont {Maruyama}}]{Singh2007}%
  \BibitemOpen
  \bibfield  {author} {\bibinfo {author} {\bibfnamefont {S.~K.}\ \bibnamefont {Singh}}, \bibinfo {author} {\bibfnamefont {H.}~\bibnamefont {Ishiwara}}, \bibinfo {author} {\bibfnamefont {K.}~\bibnamefont {Sato}}, \ and\ \bibinfo {author} {\bibfnamefont {K.}~\bibnamefont {Maruyama}},\ }\bibfield  {title} {\enquote {\bibinfo {title} {{Microstructure and frequency dependent electrical properties of Mn-substituted \ch{BiFeO3} thin films}},}\ }\href {\doibase 10.1063/1.2812594} {\bibfield  {journal} {\bibinfo  {journal} {Journal of Applied Physics}\ }\textbf {\bibinfo {volume} {102}},\ \bibinfo {pages} {094109} (\bibinfo {year} {2007})}\BibitemShut {NoStop}%
\bibitem [{\citenamefont {Kawae}\ \emph {et~al.}(2009)\citenamefont {Kawae}, \citenamefont {Terauchi}, \citenamefont {Tsuda}, \citenamefont {Kumeda},\ and\ \citenamefont {Morimoto}}]{Kawae2009ImprovedFilms}%
  \BibitemOpen
  \bibfield  {author} {\bibinfo {author} {\bibfnamefont {T.}~\bibnamefont {Kawae}}, \bibinfo {author} {\bibfnamefont {Y.}~\bibnamefont {Terauchi}}, \bibinfo {author} {\bibfnamefont {H.}~\bibnamefont {Tsuda}}, \bibinfo {author} {\bibfnamefont {M.}~\bibnamefont {Kumeda}}, \ and\ \bibinfo {author} {\bibfnamefont {A.}~\bibnamefont {Morimoto}},\ }\bibfield  {title} {\enquote {\bibinfo {title} {{Improved leakage and ferroelectric properties of Mn and Ti codoped \ch{BiFeO3} thin films}},}\ }\href {\doibase 10.1063/1.3098408} {\bibfield  {journal} {\bibinfo  {journal} {Applied Physics Letters}\ }\textbf {\bibinfo {volume} {94}},\ \bibinfo {pages} {112904} (\bibinfo {year} {2009})}\BibitemShut {NoStop}%
\bibitem [{\citenamefont {Raghavan}, \citenamefont {Kim},\ and\ \citenamefont {Kim}(2014)}]{Raghavan2014EffectsFilms}%
  \BibitemOpen
  \bibfield  {author} {\bibinfo {author} {\bibfnamefont {C.~M.}\ \bibnamefont {Raghavan}}, \bibinfo {author} {\bibfnamefont {J.~W.}\ \bibnamefont {Kim}}, \ and\ \bibinfo {author} {\bibfnamefont {S.~S.}\ \bibnamefont {Kim}},\ }\bibfield  {title} {\enquote {\bibinfo {title} {{Effects of Ho and Ti doping on structural and electrical properties of \ch{BiFeO3} thin films}},}\ }\href {\doibase 10.1111/jace.12641} {\bibfield  {journal} {\bibinfo  {journal} {Journal of the American Ceramic Society}\ }\textbf {\bibinfo {volume} {97}},\ \bibinfo {pages} {235--240} (\bibinfo {year} {2014})}\BibitemShut {NoStop}%
\bibitem [{\citenamefont {Bl\'azquez~Mart\'inez}\ \emph {et~al.}(2023{\natexlab{b}})\citenamefont {Bl\'azquez~Mart\'inez}, \citenamefont {Grysan}, \citenamefont {Girod}, \citenamefont {Kovacova}, \citenamefont {Glin\v{s}ek},\ and\ \citenamefont {Granzow}}]{BlazquezMartinez2023-Stress}%
  \BibitemOpen
  \bibfield  {author} {\bibinfo {author} {\bibfnamefont {A.}~\bibnamefont {Bl\'azquez~Mart\'inez}}, \bibinfo {author} {\bibfnamefont {P.}~\bibnamefont {Grysan}}, \bibinfo {author} {\bibfnamefont {S.}~\bibnamefont {Girod}}, \bibinfo {author} {\bibfnamefont {V.}~\bibnamefont {Kovacova}}, \bibinfo {author} {\bibfnamefont {S.}~\bibnamefont {Glin\v{s}ek}}, \ and\ \bibinfo {author} {\bibfnamefont {T.}~\bibnamefont {Granzow}},\ }\bibfield  {title} {\enquote {\bibinfo {title} {Stress-tuning the bulk photovoltaic response in polycrystalline bismuth ferrite films},}\ }\href {\doibase 10.1063/5.0136800} {\bibfield  {journal} {\bibinfo  {journal} {Applied Physics Letters}\ }\textbf {\bibinfo {volume} {122}} (\bibinfo {year} {2023}{\natexlab{b}}),\ 10.1063/5.0136800}\BibitemShut {NoStop}%
\bibitem [{\citenamefont {{Bl\'azquez Mart\'inez}}\ \emph {et~al.}(2022)\citenamefont {{Bl\'azquez Mart\'inez}}, \citenamefont {Grysan}, \citenamefont {Girod}, \citenamefont {Glin\v{s}ek},\ and\ \citenamefont {Granzow}}]{BlazquezMartinezIDE2022}%
  \BibitemOpen
  \bibfield  {author} {\bibinfo {author} {\bibfnamefont {A.}~\bibnamefont {{Bl\'azquez Mart\'inez}}}, \bibinfo {author} {\bibfnamefont {P.}~\bibnamefont {Grysan}}, \bibinfo {author} {\bibfnamefont {S.}~\bibnamefont {Girod}}, \bibinfo {author} {\bibfnamefont {S.}~\bibnamefont {Glin\v{s}ek}}, \ and\ \bibinfo {author} {\bibfnamefont {T.}~\bibnamefont {Granzow}},\ }\bibfield  {title} {\enquote {\bibinfo {title} {Direct evidence for bulk photovoltaic charge transport in a ferroelectric polycrystalline film},}\ }\href {\doibase https://doi.org/10.1016/j.scriptamat.2021.114498} {\bibfield  {journal} {\bibinfo  {journal} {Scripta Materialia}\ }\textbf {\bibinfo {volume} {211}},\ \bibinfo {pages} {114498} (\bibinfo {year} {2022})}\BibitemShut {NoStop}%
\bibitem [{\citenamefont {Aruchamy}\ \emph {et~al.}(2022)\citenamefont {Aruchamy}, \citenamefont {Schenk}, \citenamefont {Girod}, \citenamefont {Glin\v{s}ek}, \citenamefont {Defay},\ and\ \citenamefont {Granzow}}]{Aruchamy2022}%
  \BibitemOpen
  \bibfield  {author} {\bibinfo {author} {\bibfnamefont {N.}~\bibnamefont {Aruchamy}}, \bibinfo {author} {\bibfnamefont {T.}~\bibnamefont {Schenk}}, \bibinfo {author} {\bibfnamefont {S.}~\bibnamefont {Girod}}, \bibinfo {author} {\bibfnamefont {S.}~\bibnamefont {Glin\v{s}ek}}, \bibinfo {author} {\bibfnamefont {E.}~\bibnamefont {Defay}}, \ and\ \bibinfo {author} {\bibfnamefont {T.}~\bibnamefont {Granzow}},\ }\bibfield  {title} {\enquote {\bibinfo {title} {Influence of substrate stress on in-plane and out-of-plane ferroelectric properties of {PZT} films},}\ }\href {\doibase 10.1063/5.0072503} {\bibfield  {journal} {\bibinfo  {journal} {Journal of Applied Physics}\ }\textbf {\bibinfo {volume} {131}} (\bibinfo {year} {2022}),\ 10.1063/5.0072503}\BibitemShut {NoStop}%
\bibitem [{\citenamefont {Zeiss}(2008)}]{zeiss_microscopy_2008}%
  \BibitemOpen
  \bibfield  {author} {\bibinfo {author} {\bibfnamefont {C.}~\bibnamefont {Zeiss}},\ }\href {http://www.fabtech.org/news/_a/microscopy_resolution_record_claimed_by_carl_zeiss/} {\enquote {\bibinfo {title} {Microscopy resolution record claimed by {Carl} {Zeiss}, (http://www.fabtech.org/news)},}\ }\bibinfo {type} {Tech. Rep.}\ (\bibinfo  {institution} {Zeiss, Carl},\ \bibinfo {year} {2008})\BibitemShut {NoStop}%
\bibitem [{\citenamefont {Herklotz}\ \emph {et~al.}(2019{\natexlab{b}})\citenamefont {Herklotz}, \citenamefont {Rus}, \citenamefont {Sohn}, \citenamefont {KC}, \citenamefont {Cooper}, \citenamefont {Guo},\ and\ \citenamefont {Ward}}]{herklotz_optical_2019}%
  \BibitemOpen
  \bibfield  {author} {\bibinfo {author} {\bibfnamefont {A.}~\bibnamefont {Herklotz}}, \bibinfo {author} {\bibfnamefont {S.~F.}\ \bibnamefont {Rus}}, \bibinfo {author} {\bibfnamefont {C.}~\bibnamefont {Sohn}}, \bibinfo {author} {\bibfnamefont {S.}~\bibnamefont {KC}}, \bibinfo {author} {\bibfnamefont {V.~R.}\ \bibnamefont {Cooper}}, \bibinfo {author} {\bibfnamefont {E.-J.}\ \bibnamefont {Guo}}, \ and\ \bibinfo {author} {\bibfnamefont {T.~Z.}\ \bibnamefont {Ward}},\ }\bibfield  {title} {\enquote {\bibinfo {title} {Optical response of {BiFeO$_3$} films subjected to uniaxial strain},}\ }\href {\doibase 10.1103/PhysRevMaterials.3.094410} {\bibfield  {journal} {\bibinfo  {journal} {Physical Review Materials}\ }\textbf {\bibinfo {volume} {3}},\ \bibinfo {pages} {094410} (\bibinfo {year} {2019}{\natexlab{b}})},\ \bibinfo {note} {publisher: American Physical Society}\BibitemShut {NoStop}%
\bibitem [{\citenamefont {Sando}\ \emph {et~al.}(2014)\citenamefont {Sando}, \citenamefont {Hermet}, \citenamefont {Allibe}, \citenamefont {Bourderionnet}, \citenamefont {Fusil}, \citenamefont {Carrétéro}, \citenamefont {Jacquet}, \citenamefont {Mage}, \citenamefont {Dolfi}, \citenamefont {Barthélémy}, \citenamefont {Ghosez},\ and\ \citenamefont {Bibes}}]{sando_linear_2014}%
  \BibitemOpen
  \bibfield  {author} {\bibinfo {author} {\bibfnamefont {D.}~\bibnamefont {Sando}}, \bibinfo {author} {\bibfnamefont {P.}~\bibnamefont {Hermet}}, \bibinfo {author} {\bibfnamefont {J.}~\bibnamefont {Allibe}}, \bibinfo {author} {\bibfnamefont {J.}~\bibnamefont {Bourderionnet}}, \bibinfo {author} {\bibfnamefont {S.}~\bibnamefont {Fusil}}, \bibinfo {author} {\bibfnamefont {C.}~\bibnamefont {Carrétéro}}, \bibinfo {author} {\bibfnamefont {E.}~\bibnamefont {Jacquet}}, \bibinfo {author} {\bibfnamefont {J.-C.}\ \bibnamefont {Mage}}, \bibinfo {author} {\bibfnamefont {D.}~\bibnamefont {Dolfi}}, \bibinfo {author} {\bibfnamefont {A.}~\bibnamefont {Barthélémy}}, \bibinfo {author} {\bibfnamefont {P.}~\bibnamefont {Ghosez}}, \ and\ \bibinfo {author} {\bibfnamefont {M.}~\bibnamefont {Bibes}},\ }\bibfield  {title} {\enquote {\bibinfo {title} {Linear electro-optic effect in multiferroic {BiFeO$_3$} thin films},}\ }\href {\doibase 10.1103/PhysRevB.89.195106} {\bibfield  {journal} {\bibinfo  {journal} {Physical Review B}\
  }\textbf {\bibinfo {volume} {89}},\ \bibinfo {pages} {195106} (\bibinfo {year} {2014})},\ \bibinfo {note} {publisher: American Physical Society}\BibitemShut {NoStop}%
\bibitem [{\citenamefont {Ziegler}(2004)}]{ziegler_srim-2003_2004}%
  \BibitemOpen
  \bibfield  {author} {\bibinfo {author} {\bibfnamefont {J.~F.}\ \bibnamefont {Ziegler}},\ }\bibfield  {title} {\enquote {\bibinfo {title} {{SRIM}-2003},}\ }\href {\doibase 10.1016/j.nimb.2004.01.208} {\bibfield  {journal} {\bibinfo  {journal} {Nuclear Instruments and Methods in Physics Research Section B: Beam Interactions with Materials and Atoms}\ }\bibinfo {series} {Proceedings of the {Sixteenth} {International} {Conference} on {Ion} {Beam} {Analysis}},\ \textbf {\bibinfo {volume} {219-220}},\ \bibinfo {pages} {1027--1036} (\bibinfo {year} {2004})}\BibitemShut {NoStop}%
\bibitem [{\citenamefont {Guo}\ \emph {et~al.}(2018)\citenamefont {Guo}, \citenamefont {Wang}, \citenamefont {Yuan}, \citenamefont {He}, \citenamefont {Lu}, \citenamefont {Chen}, \citenamefont {Yang}, \citenamefont {Wang}, \citenamefont {Erni}, \citenamefont {Rossell}, \citenamefont {Gopalan}, \citenamefont {Xiang}, \citenamefont {Tokura},\ and\ \citenamefont {Yu}}]{guo_strain-induced_2018}%
  \BibitemOpen
  \bibfield  {author} {\bibinfo {author} {\bibfnamefont {J.~W.}\ \bibnamefont {Guo}}, \bibinfo {author} {\bibfnamefont {P.~S.}\ \bibnamefont {Wang}}, \bibinfo {author} {\bibfnamefont {Y.}~\bibnamefont {Yuan}}, \bibinfo {author} {\bibfnamefont {Q.}~\bibnamefont {He}}, \bibinfo {author} {\bibfnamefont {J.~L.}\ \bibnamefont {Lu}}, \bibinfo {author} {\bibfnamefont {T.~Z.}\ \bibnamefont {Chen}}, \bibinfo {author} {\bibfnamefont {S.~Z.}\ \bibnamefont {Yang}}, \bibinfo {author} {\bibfnamefont {Y.~J.}\ \bibnamefont {Wang}}, \bibinfo {author} {\bibfnamefont {R.}~\bibnamefont {Erni}}, \bibinfo {author} {\bibfnamefont {M.~D.}\ \bibnamefont {Rossell}}, \bibinfo {author} {\bibfnamefont {V.}~\bibnamefont {Gopalan}}, \bibinfo {author} {\bibfnamefont {H.~J.}\ \bibnamefont {Xiang}}, \bibinfo {author} {\bibfnamefont {Y.}~\bibnamefont {Tokura}}, \ and\ \bibinfo {author} {\bibfnamefont {P.}~\bibnamefont {Yu}},\ }\bibfield  {title} {\enquote {\bibinfo {title} {Strain-induced ferroelectricity and spin-lattice coupling in
  \ch{SrMnO3} thin films},}\ }\href {\doibase 10.1103/PhysRevB.97.235135} {\bibfield  {journal} {\bibinfo  {journal} {Physical Review B}\ }\textbf {\bibinfo {volume} {97}},\ \bibinfo {pages} {235135} (\bibinfo {year} {2018})}\BibitemShut {NoStop}%
\bibitem [{\citenamefont {Daumont}\ \emph {et~al.}(2012)\citenamefont {Daumont}, \citenamefont {Ren}, \citenamefont {Infante}, \citenamefont {Lisenkov}, \citenamefont {Allibe}, \citenamefont {Carr{\'e}t{\'e}ro}, \citenamefont {Fusil}, \citenamefont {Jacquet}, \citenamefont {Bouvet}, \citenamefont {Bouamrane} \emph {et~al.}}]{daumont2012strain}%
  \BibitemOpen
  \bibfield  {author} {\bibinfo {author} {\bibfnamefont {C.}~\bibnamefont {Daumont}}, \bibinfo {author} {\bibfnamefont {W.}~\bibnamefont {Ren}}, \bibinfo {author} {\bibfnamefont {I.}~\bibnamefont {Infante}}, \bibinfo {author} {\bibfnamefont {S.}~\bibnamefont {Lisenkov}}, \bibinfo {author} {\bibfnamefont {J.}~\bibnamefont {Allibe}}, \bibinfo {author} {\bibfnamefont {C.}~\bibnamefont {Carr{\'e}t{\'e}ro}}, \bibinfo {author} {\bibfnamefont {S.}~\bibnamefont {Fusil}}, \bibinfo {author} {\bibfnamefont {E.}~\bibnamefont {Jacquet}}, \bibinfo {author} {\bibfnamefont {T.}~\bibnamefont {Bouvet}}, \bibinfo {author} {\bibfnamefont {F.}~\bibnamefont {Bouamrane}},  \emph {et~al.},\ }\bibfield  {title} {\enquote {\bibinfo {title} {Strain dependence of polarization and piezoelectric response in epitaxial \ch{BiFeO3} thin films},}\ }\href@noop {} {\bibfield  {journal} {\bibinfo  {journal} {Journal of Physics: Condensed Matter}\ }\textbf {\bibinfo {volume} {24}},\ \bibinfo {pages} {162202} (\bibinfo {year} {2012})}\BibitemShut
  {NoStop}%
\bibitem [{\citenamefont {Biegalski}\ \emph {et~al.}(2011)\citenamefont {Biegalski}, \citenamefont {Kim}, \citenamefont {Choudhury}, \citenamefont {Chen}, \citenamefont {Christen},\ and\ \citenamefont {D{\"o}rr}}]{biegalski2011strong}%
  \BibitemOpen
  \bibfield  {author} {\bibinfo {author} {\bibfnamefont {M.~D.}\ \bibnamefont {Biegalski}}, \bibinfo {author} {\bibfnamefont {D.~H.}\ \bibnamefont {Kim}}, \bibinfo {author} {\bibfnamefont {S.}~\bibnamefont {Choudhury}}, \bibinfo {author} {\bibfnamefont {L.}~\bibnamefont {Chen}}, \bibinfo {author} {\bibfnamefont {H.}~\bibnamefont {Christen}}, \ and\ \bibinfo {author} {\bibfnamefont {K.}~\bibnamefont {D{\"o}rr}},\ }\bibfield  {title} {\enquote {\bibinfo {title} {Strong strain dependence of ferroelectric coercivity in a \ch{BiFeO3} film},}\ }\href@noop {} {\bibfield  {journal} {\bibinfo  {journal} {Applied Physics Letters}\ }\textbf {\bibinfo {volume} {98}} (\bibinfo {year} {2011})}\BibitemShut {NoStop}%
\bibitem [{\citenamefont {Jang}\ \emph {et~al.}(2008)\citenamefont {Jang}, \citenamefont {Baek}, \citenamefont {Ortiz}, \citenamefont {Folkman}, \citenamefont {Das}, \citenamefont {Chu}, \citenamefont {Shafer}, \citenamefont {Zhang}, \citenamefont {Choudhury}, \citenamefont {Vaithyanathan} \emph {et~al.}}]{jang2008strain}%
  \BibitemOpen
  \bibfield  {author} {\bibinfo {author} {\bibfnamefont {H.}~\bibnamefont {Jang}}, \bibinfo {author} {\bibfnamefont {S.}~\bibnamefont {Baek}}, \bibinfo {author} {\bibfnamefont {D.}~\bibnamefont {Ortiz}}, \bibinfo {author} {\bibfnamefont {C.}~\bibnamefont {Folkman}}, \bibinfo {author} {\bibfnamefont {R.}~\bibnamefont {Das}}, \bibinfo {author} {\bibfnamefont {Y.}~\bibnamefont {Chu}}, \bibinfo {author} {\bibfnamefont {P.}~\bibnamefont {Shafer}}, \bibinfo {author} {\bibfnamefont {J.}~\bibnamefont {Zhang}}, \bibinfo {author} {\bibfnamefont {S.}~\bibnamefont {Choudhury}}, \bibinfo {author} {\bibfnamefont {V.}~\bibnamefont {Vaithyanathan}},  \emph {et~al.},\ }\bibfield  {title} {\enquote {\bibinfo {title} {Strain-induced polarization rotation in epitaxial (001) \ch{BiFeO3} thin films},}\ }\href@noop {} {\bibfield  {journal} {\bibinfo  {journal} {Physical review letters}\ }\textbf {\bibinfo {volume} {101}},\ \bibinfo {pages} {107602} (\bibinfo {year} {2008})}\BibitemShut {NoStop}%
\bibitem [{\citenamefont {Saremi}\ \emph {et~al.}(2018{\natexlab{a}})\citenamefont {Saremi}, \citenamefont {Xu}, \citenamefont {Dedon}, \citenamefont {Gao}, \citenamefont {Ghosh}, \citenamefont {Dasgupta},\ and\ \citenamefont {Martin}}]{Saremi2018}%
  \BibitemOpen
  \bibfield  {author} {\bibinfo {author} {\bibfnamefont {S.}~\bibnamefont {Saremi}}, \bibinfo {author} {\bibfnamefont {R.}~\bibnamefont {Xu}}, \bibinfo {author} {\bibfnamefont {L.~R.}\ \bibnamefont {Dedon}}, \bibinfo {author} {\bibfnamefont {R.}~\bibnamefont {Gao}}, \bibinfo {author} {\bibfnamefont {A.}~\bibnamefont {Ghosh}}, \bibinfo {author} {\bibfnamefont {A.}~\bibnamefont {Dasgupta}}, \ and\ \bibinfo {author} {\bibfnamefont {L.~W.}\ \bibnamefont {Martin}},\ }\bibfield  {title} {\enquote {\bibinfo {title} {Electronic transport and ferroelectric switching in ion‐bombarded, defect‐engineered \ch{BiFeO3} thin films},}\ }\href {\doibase 10.1002/admi.201700991} {\bibfield  {journal} {\bibinfo  {journal} {Advanced Materials Interfaces}\ }\textbf {\bibinfo {volume} {5}} (\bibinfo {year} {2018}{\natexlab{a}}),\ 10.1002/admi.201700991}\BibitemShut {NoStop}%
\bibitem [{\citenamefont {Saremi}\ \emph {et~al.}(2016)\citenamefont {Saremi}, \citenamefont {Xu}, \citenamefont {Dedon}, \citenamefont {Mundy}, \citenamefont {Hsu}, \citenamefont {Chen}, \citenamefont {Damodaran}, \citenamefont {Chapman}, \citenamefont {Evans},\ and\ \citenamefont {Martin}}]{Saremi2016}%
  \BibitemOpen
  \bibfield  {author} {\bibinfo {author} {\bibfnamefont {S.}~\bibnamefont {Saremi}}, \bibinfo {author} {\bibfnamefont {R.}~\bibnamefont {Xu}}, \bibinfo {author} {\bibfnamefont {L.~R.}\ \bibnamefont {Dedon}}, \bibinfo {author} {\bibfnamefont {J.~A.}\ \bibnamefont {Mundy}}, \bibinfo {author} {\bibfnamefont {S.}~\bibnamefont {Hsu}}, \bibinfo {author} {\bibfnamefont {Z.}~\bibnamefont {Chen}}, \bibinfo {author} {\bibfnamefont {A.~R.}\ \bibnamefont {Damodaran}}, \bibinfo {author} {\bibfnamefont {S.~P.}\ \bibnamefont {Chapman}}, \bibinfo {author} {\bibfnamefont {J.~T.}\ \bibnamefont {Evans}}, \ and\ \bibinfo {author} {\bibfnamefont {L.~W.}\ \bibnamefont {Martin}},\ }\bibfield  {title} {\enquote {\bibinfo {title} {Enhanced electrical resistivity and properties via ion bombardment of ferroelectric thin films},}\ }\href {\doibase 10.1002/adma.201603968} {\bibfield  {journal} {\bibinfo  {journal} {Advanced Materials}\ }\textbf {\bibinfo {volume} {28}},\ \bibinfo {pages} {10750--10756} (\bibinfo {year}
  {2016})}\BibitemShut {NoStop}%
\bibitem [{\citenamefont {Saremi}\ \emph {et~al.}(2018{\natexlab{b}})\citenamefont {Saremi}, \citenamefont {Xu}, \citenamefont {Allen}, \citenamefont {Maher}, \citenamefont {Agar}, \citenamefont {Gao}, \citenamefont {Hosemann},\ and\ \citenamefont {Martin}}]{saremi_local_2018}%
  \BibitemOpen
  \bibfield  {author} {\bibinfo {author} {\bibfnamefont {S.}~\bibnamefont {Saremi}}, \bibinfo {author} {\bibfnamefont {R.}~\bibnamefont {Xu}}, \bibinfo {author} {\bibfnamefont {F.~I.}\ \bibnamefont {Allen}}, \bibinfo {author} {\bibfnamefont {J.}~\bibnamefont {Maher}}, \bibinfo {author} {\bibfnamefont {J.~C.}\ \bibnamefont {Agar}}, \bibinfo {author} {\bibfnamefont {R.}~\bibnamefont {Gao}}, \bibinfo {author} {\bibfnamefont {P.}~\bibnamefont {Hosemann}}, \ and\ \bibinfo {author} {\bibfnamefont {L.~W.}\ \bibnamefont {Martin}},\ }\bibfield  {title} {\enquote {\bibinfo {title} {Local control of defects and switching properties in ferroelectric thin films},}\ }\href {\doibase 10.1103/PhysRevMaterials.2.084414} {\bibfield  {journal} {\bibinfo  {journal} {Physical Review Materials}\ }\textbf {\bibinfo {volume} {2}},\ \bibinfo {pages} {084414} (\bibinfo {year} {2018}{\natexlab{b}})},\ \bibinfo {note} {publisher: American Physical Society}\BibitemShut {NoStop}%
\bibitem [{\citenamefont {Kim}\ \emph {et~al.}(2020)\citenamefont {Kim}, \citenamefont {Saremi}, \citenamefont {Acharya}, \citenamefont {Velarde}, \citenamefont {Parsonnet}, \citenamefont {Donahue}, \citenamefont {Qualls}, \citenamefont {Garcia},\ and\ \citenamefont {Martin}}]{Kim2020}%
  \BibitemOpen
  \bibfield  {author} {\bibinfo {author} {\bibfnamefont {J.}~\bibnamefont {Kim}}, \bibinfo {author} {\bibfnamefont {S.}~\bibnamefont {Saremi}}, \bibinfo {author} {\bibfnamefont {M.}~\bibnamefont {Acharya}}, \bibinfo {author} {\bibfnamefont {G.}~\bibnamefont {Velarde}}, \bibinfo {author} {\bibfnamefont {E.}~\bibnamefont {Parsonnet}}, \bibinfo {author} {\bibfnamefont {P.}~\bibnamefont {Donahue}}, \bibinfo {author} {\bibfnamefont {A.}~\bibnamefont {Qualls}}, \bibinfo {author} {\bibfnamefont {D.}~\bibnamefont {Garcia}}, \ and\ \bibinfo {author} {\bibfnamefont {L.~W.}\ \bibnamefont {Martin}},\ }\bibfield  {title} {\enquote {\bibinfo {title} {Ultrahigh capacitive energy density in ion-bombarded relaxor ferroelectric films},}\ }\href {\doibase 10.1126/science.abb0631} {\bibfield  {journal} {\bibinfo  {journal} {Science}\ }\textbf {\bibinfo {volume} {369}},\ \bibinfo {pages} {81--84} (\bibinfo {year} {2020})}\BibitemShut {NoStop}%
\end{thebibliography}%

\end{document}